\documentclass[12pt, titlepage]{article}

\usepackage{alltt}
\usepackage{graphicx, amsmath, amssymb, natbib, setspace}
\usepackage{float}
\usepackage{booktabs,array}
\usepackage[caption = false]{subfig}
\usepackage{cancel}
\usepackage{xcolor}

\usepackage{mymacros}
\usepackage{lineno}
\usepackage{caption}
 
\usepackage{makecell}
\setcellgapes{5pt}

\IfFileExists{upquote.sty}{\usepackage{upquote}}{}
\begin{document}
\singlespacing

\titlepage
\title {Marginal Inference for Hierarchical Generalized Linear Mixed Models with Patterned Covariance Matrices Using the Laplace Approximation}
\author{Jay M.\ Ver Hoef$^{1}$, Eryn Blagg$^{2}$, Michael Dumelle$^{3}$, Philip M.\ Dixon$^{2}$,  \\
 Dale L. Zimmerman$^{4}$, Paul Conn$^{1}$ \\
\hrulefill \\ 
$^{1}$Marine Mammal Laboratory \\
NOAA-NMFS Alaska Fisheries Science Center\\
7600 Sand Point Way NE, Seattle, WA 98115\\
E-mail: jay.verhoef@noaa.gov\\
\hrulefill \\
$^{2}$ Department of Statistics, Iowa State University, Ames, Iowa \\
$^{3}$ United States Environmental Protection Agency, \\
	200 SW 35th St, Corvallis, Oregon \\
$^{4}$ Department of Statistics and Actuarial Science, \\
	The University of Iowa, Iowa City, Iowa \\
\hrulefill \\
}

\maketitle

\doublespacing


\begin{abstract}
Using a hierarchical construction, we develop methods for a wide and flexible class of models by taking a fully parametric approach to generalized linear mixed models with complex covariance dependence. The Laplace approximation is used to marginally estimate covariance parameters while integrating out all fixed and latent random effects. The Laplace approximation relies on Newton-Raphson updates, which also leads to predictions for the latent random effects.  We develop methodology for complete marginal inference, from estimating covariance parameters and fixed effects to making predictions for unobserved data, for any patterned covariance matrix in the hierarchical generalized linear mixed models framework. The marginal likelihood is developed for six distributions that are often used for binary, count, and positive continuous data, and our framework is easily extended to other distributions.  The methods are illustrated with simulations from stochastic processes with known parameters, and their efficacy in terms of bias and interval coverage is shown through simulation experiments. Examples with binary and proportional data on election results, count data for marine mammals, and positive-continuous data on heavy metal concentration in the environment are used to illustrate all six distributions with a variety of patterned covariance structures that include spatial models (e.g., geostatistical and areal models), time series models (e.g., first-order autoregressive models), and mixtures with typical random intercepts based on grouping.
\end{abstract}


\newpage
\section{INTRODUCTION}

The classical linear model relies on a normal distribution that has continuous support on the real line, but many data are binary, counts, or positive continuous.  Such data can be transformed to stabilize variances and create empirical distributions that are ``near normal,'' allowing the use of classical linear models \citep[e.g.,][ p. 288]{SnedecorEtAl1980StatisticalMethodsSeventh}. For example, a square root transformation can be used for count data.  However, \citet{NelderEtAl1972GeneralizedLinearModels370} introduced the generalized linear model \citep[GLM,][]{McCullaghEtAl1989GeneralizedLinearModels} as a natural extension to linear models, such as the Poisson distribution for counts, the Bernoulli distribution for binary data, etc., which have become very popular and generally preferred to data transformations \citep[e.g.,][]{warton_arcsine_2011}.  GLMs can be extended by introducing latent random effects via a linear mixed model to create a class of generalized linear mixed models \citep[GLMMs,][]{breslow_approximate_1993}.  These latent random effects are usually assumed to be independent and identically distributed normal variables \citep{zeger_generalized_1991}, however it is also possible for the latent random effects to be temporally autocorrelated \citep[e.g.,][]{stiratelli_random-effects_1984, zeger_models_1988}, spatially autocorrelated \citep[e.g,][]{clayton_empirical_1987, GotwayEtAl1997GeneralizedLinearModel157, DiggleEtAl1998ModelbasedGeostatisticsDisc299}, or both \citep[][p. 380]{CressieEtAl2011StatisticsSpatiotemporalData}.  A unifying framework for this literature is available through a hierarchical generalized linear mixed model \citep[HGLMM,][]{lee_hierarchical_1996}.


\subsection{Hierarchical Generalized Linear Mixed Models}

Most GLMs are motivated by the exponential family of distributions \citep[e.g.,][]{fisher_two_1934, lehmann_theory_2006}.  However, most common software packages use iteratively-reweighted least-squares \citep{wedderburn_quasi-likelihood_1974} to fit GLMs through their first two moments (mean and variance functions). This procedure allows a flexible class of models that can be fit though a single inferential procedure (quasi-likelihood), which is desirable because some GLMs like the quasi-Poisson have no true likelihood.

By contrast, we will take a fully parametric approach to create covariance dependence for GLMMs through a hierarchical construction.  We will use the notation $[\by|\bmu]$ to denote any joint probability density function of the vector of random variables $\by$ conditional on a vector of parameters, or other fixed variables, $\bmu$. In some cases, we may want to model multiple vectors of responses (as with time series applications) or to condition on more than one set of parameters, in which case an expression might look more like  $[\by_{1},\by_{2}, \ldots, \by_{k} | \bmu_{1}, \bmu_{2}, \ldots, \bmu_{\ell},\bphi]$. As a simple example, consider a joint probability mass function that consists of the product of independent negative binomial distributions, which can be parameterized with a mean vector and a common extra parameter that allows for overdispersion. We could write this as $[\by|\bmu,\phi]$, where the $\bmu$ represents the mean vector and $\phi$ is the overdispersion parameter.

For the hierarchical construction of the generalized linear mixed models that we consider in this article, we allow the mean to vary by other random variables $\bw$, and we condition on these random variables, $[\by|\bg^{-1}(\bw),\bphi]$, through the multivariate mean function $\textrm{E}(\by) = \bg^{-1}(\mathbf{w})$. For the Poisson distribution, each element of $\bg(\cdot)$ is often the log function, and in general $\bg(\cdot)$ is called the (multivariate) link function \citep{McCullaghEtAl1989GeneralizedLinearModels}.  Link functions are monotonic so that each element of $\bg^{-1}(\cdot)$ is one-to-one with each element of $g(\cdot)$. Recall that the mean of a Poisson distribution must be positive, and if $\bg(\cdot)$ has each element as the log function, then $\bg^{-1}(\cdot)$ has each element as the exponential function so $\bmu = \bg^{-1}(\bw)$ always has positive elements, which allows each element of $\bw$ to be unconstrained.

We will only consider models where $\bw$ is $n \times 1$ and has a multivariate normal distribution that is constructed through the linear mixed model,
\begin{linenomath*}
\begin{equation} \label{eq:lm_eta}
\bw = \bX\bbeta + \sum_{k=1}^{q}\bZ_{k}\br_{k} + \bepsilon,
\end{equation}
\end{linenomath*}
where $\bX$ is an $n \times p$ full rank design matrix of explanatory variables, $\bbeta$ is a $p \times 1$ parameter vector of fixed effects, $\bZ_{k}$ is a design matrix for the $k$th random effect vector $\br_{k}$, and $\bepsilon$ represents independent and identically-distributed Gaussian error.  We assume that $\textrm{E}(\br_{k}) = \bzero$ for all $k$, $\textrm{E}(\bepsilon) = \bzero$, $\var(\br_{k}) = \bV_{k}$, $\cov(\br_{j}, \br_{k}) = \bzero$ when $j \ne k$, and $\var(\bepsilon) = \sigma^{2}_{0}\bI$. We use the notation $[\bw|\bX,\bbeta,\{\bZ_{k}\},\{\bV_{k}\},\btheta]$ to indicate the probability density function $\bw \sim N(\bX\bbeta,\bSigma_{\btheta})$ where
\begin{linenomath*} 
$$
\bSigma_{\btheta} = \sum_{k}\bZ_{k}\bV_{k}\bZ_{k}\upp + \sigma^{2}_{0}\bI.
$$
\end{linenomath*}
The covariance matrices $\{\bV_{k}\}$ for $k = 1,\ldots,q$ can have additional covariance parameters beyond $\sigma^{2}_{0}$, all of which are contained in the vector $\btheta$.  We will give more specific details on $\bSigma_{\btheta}$ later.

For the fully parametric, hierarchical models, a very general model can be constructed hierarchically as,
\begin{linenomath*}
\begin{equation} \label{eq:hglmm}
[\by,\bw|\bphi,\bX,\bbeta,\bSigma_{\btheta}] = [\by|\bg^{-1}(\bw),\bphi][\bw|\bX, \bbeta,\bSigma_{\btheta}],
\end{equation}
\end{linenomath*}
where \citet{berliner_hierarchical_1996} called $[\by|\bg^{-1}(\bw),\bphi]$ the \textit{data model} and $[\bw|\bX, \bbeta,\bSigma_{\btheta}]$ the \textit{process model}. As a concrete example, suppose that $[\by|\exp(\bw)]$ is Poisson, and $[\bw|\bX,\bbeta,\bSigma_{\btheta}]$ is multivariate normal, then the joint likelihood is
\begin{linenomath*}
$$
[\by,\bw|\bX,\bbeta,\bSigma_{\btheta}] = \left(\prod_{i=1}^{n}\frac{\exp(w_{i})^{y_{i}}\exp(-\exp(w_i))}{y_{i}!}\right)\frac{\exp\left(-0.5(\bw - \bX\bbeta)\upp\bSigma_{\btheta}^{-1}(\bw - \bX\bbeta)\right)}{(2\pi)^{n/2}|\bSigma_{\btheta}|^{1/2}},
$$
\end{linenomath*}
and note the use of $\exp(w_{i})$ for $\textrm{E}(y_{i})$.


\subsection{Patterned Covariance Matrices} \label{sec:patCovMat}

To construct a likelihood for \eqref{eq:hglmm} we will need parametric models for $\bSigma_{\btheta}$ in \eqref{eq:lm_eta}.  There are few constraints here, and any valid covariance model for $\bSigma_{\btheta}$ is possible. For example, $\bSigma_{\btheta}$ may be constructed from typical mixed models where $\bZ_{k}$ contains indicator variables for random intercepts, or explanatory variables for random slopes, and where $\bV_{k} = \sigma_{k}^{2}\bI$, and then $\bSigma_{\btheta} = \sum_{k}\sigma^{2}_{k}\bZ_{k}\bZ_{k}\upp + \sigma^{2}_{0}\bI$. We can also consider time series models \citep[e.g.,][]{Hamilton1994TimeSeriesAnalysis}.  For example, for a first-order autoregressive (AR1) model with $i$ and $j$ being integers, let $q = 1$ and $\bZ_{1} = \bI$, then $\bV_{1}$ has as its $i,j$th entry $\sigma_{1}^{2}\rho^{|i-j|}/(1 - \rho^{2})$, where $0 < \sigma^{2}_{1} $ and $0 \le \rho < 1$. Similarly, we can have geostatistical models \citep[e.g.,][]{ChilesEtAl1999GeostatisticsModelingSpatial}, such as the exponential autocovariance model, where $q = 1$, $\bZ_{1} = \bI$, and the $i,j$th element of $\bV_{1}$ is $\sigma^{2}_{1}\exp(-\delta_{i,j}/\rho)$ where $\delta_{i,j}$ is Euclidean distance between the $i$th and $j$th locations, $0 < \sigma^{2}_{1}$, and $\rho > 0$.  Other spatial covariance types include the conditional autoregressive \citep[CAR, ][]{Besag1974SpatialInteractionStatistical192,Cressie1993StatisticsSpatialData} and simultaneous autoregressive models \citep[SAR, ][]{Whittle1954stationaryprocessesplane434,VerHoefEtAl2018Spatialautoregressivemodels36}, moving average models in time series \citep[e.g.,][]{Hamilton1994TimeSeriesAnalysis} and spatial statistics \citep{haining_moving_1978}, spatio-temporal models \citep[][p. 380]{CressieEtAl2011StatisticsSpatiotemporalData}, and models on non-Euclidean topologies such as a sphere \citep[e.g., the earth,][]{huang_validity_2011,gneiting_strictly_2013}, and networks such as roads \citep{VerHoef2018KrigingModelsLinear1600} and streams \citep{VerHoefEtAl2010movingaverageapproach6}. Because a covariance matrix can be constructed by summing covariance matrices as variance components, mixtures of all models mentioned above can create a rich set of patterned covariance matrices for modeling dependent structures.  In what follows, we develop inference based on any valid covariance matrix. 


\subsection{Inference for HGLMMs}

The combination of the data model, $[\by|\bg^{-1}(\bw),\bphi]$, where any distribution could be used that matches the type of data, and the process model, $[\bw|\bX, \bbeta,\bSigma_{\btheta}]$, that can allow for any patterned covariance matrix, provides a hierarchical construction \eqref{eq:hglmm} that is a very rich and flexible class of models.  This class of models is not new.

There are two broad methods of analysis.  The most obvious method is Bayesian and computes the posterior distribution of all latent variables and parameters.  Due to intractable integrals, this is usually achieved with Markov chain Monte Carlo (MCMC) methods \citep{GelfandEtAl1990Samplingbasedapproachescalculating398, GilksEtAl1996IntroducingMarkovchain1}, of which there are now many varieties.  Bayesian hierarchical models in our context have been extremely popular, beginning with spatial statistics \citep[e.g.][]{clayton_empirical_1987}, clustered data \citep[e.g.][]{zeger_generalized_1991}, time series \citep[e.g.][]{berliner_hierarchical_1996}, and longitudinal data \citep{kleinman_semi-parametric_1998} among others, and have been aided by the introduction of the \texttt{WinBUGS} software \citep{lunn_winbugs_2000}. 

A second approach attempts to estimate covariance parameters and fixed effects marginally while integrating out over all latent random effects.  This can also be done using MCMC methods as a numerical integrator \citep[e.g.,][]{zhang_estimation_2002, christensen_monte_2004}, but it can be quite slow, so a more popular and deterministic method uses a Laplace approximation \citep{tierney_accurate_1986}.  In particular, \citet{rue_approximate_2009} proposed integrated nested Laplace approximation (INLA) as approximate Bayesian inference when using Gaussian Markov random fields.  They exploit the marginal specification of conditional autoregressive models for computational gains and use several first-order Laplace approximations to estimate fixed effects. \citet{wood_simplified_2020} provided further computational gains with dense covariance matrices. Automatic differentiation is used in software \texttt{glmmTMB} \citep{brooks_glmmtmb_2017} as a general computational approach to fitting these models.  Our development builds primarily on \citet{evangelou_estimation_2011} and \citet{ bonat_practical_2016}, who use a second-order Laplace approximation with geostatistical covariance structures for binary and count data.  The more general formulation is given by \citet{ bonat_practical_2016}, and we will point out differences from our development and that of \citet{ bonat_practical_2016} in our Methods section. 

Despite the relatively long history of this subject, there is no unified framework for the case where covariance matrices are patterned by spatial or temporal correlations or subject to other commonly used random effects.  Our general goal is to provide complete marginal inference, from estimating covariance parameters and fixed effects, to making predictions for unobserved data, for any patterned covariance matrix in the HGLMM framework.  In particular, our goals are to: 1) find marginal maximum likelihood and restricted maximum likelihood estimates for covariance parameters $\btheta$ and $\bphi$, 2) predict the latent values of $\mathbf{w}$, 3) estimate fixed effects $\bbeta$, and 4) make predictions of new values of the process that generated $\mathbf{w}$ at unsampled times or locations.

The rest of this paper is organized as follows. In Section 2, we use the Laplace approximation to develop marginal maximum likelihood estimates for $\btheta$ and $\bphi$ using Newton-Raphson updates, which also leads to predictions of $\mathbf{w}$.  From the predictions of $\mathbf{w}$ we develop estimators of $\bbeta$ with proper confidence intervals and prediction of new values of the process generating $\mathbf{w}$ with proper prediction intervals.  In Section 3, we conduct simulations to illustrate all methods and validate the earlier development.  Section 4 provides three separate examples using real data sets to further illustrate the methods.  We conclude with some discussion in Section 5.


\section{Methods}

When considering the hierarchical model formulation of the HGLMMs, we would like to marginalize the distribution $[\mathbf{w}, \mathbf{y}|\bphi,\bX,\bbeta,\bSigma_{\btheta}] = [\mathbf{y}|\bg^{-1}(\mathbf{w}),\bphi][\mathbf{w}|\bX,\bbeta,\bSigma_{\btheta}]$ over $\mathbf{w}$ and be free of $\bbeta$ as well to obtain a distribution of only the data and variance/covariance parameters. The Laplace method helps achieve that.


\subsection{Laplace for HGLMMs}

First, consider integrating over $\bbeta$ as well as $\mathbf{w}$,
\begin{linenomath*}
$$
	[\mathbf{y}|\bphi,\btheta] = \int_{\mathbf{w} \in \mathbb{R}^n} \int_{\bbeta\in \mathbb{R}^p} [\mathbf{w}, \mathbf{y}|\bphi,\bX,\bbeta,\bSigma_{\btheta}] d\bbeta d\mathbf{w} =
		\int_\mathbf{w}  [\mathbf{y}|\bg^{-1}(\mathbf{w}),\bphi] \int_{\bbeta} [\mathbf{w}|\bX,\bbeta,\bSigma_{\btheta}] d\bbeta d\mathbf{w}.
$$
\end{linenomath*}
When $[\mathbf{w}|\bX,\bbeta,\bSigma_{\btheta}]$ is Gaussian, $\int_{\bbeta} [\mathbf{w}|\bX,\bbeta,\bSigma_{\btheta}] d\bbeta$ is the likelihood for restricted (also known as residual) maximum likelihood estimation (REML, see Appendix). Note that REML was originally derived as the likelihood of a set of $n - p$ independent linear combinations of the observations known as error contrasts \citep{PattersonEtAl1971Recoveryinterblockinformation545,PattersonEtAl1974Maximumlikelihoodestimation197}, and there is little literature on its derivation from integration.  Alternatively, consider $[\mathbf{w}|\bX,\bbeta,\bSigma_{\btheta}]$ where $\bbeta$ has been replaced by its conditional (on $\mathbf{w}$) maximum likelihood (ML) estimator, $\hat{\bbeta} = (\mathbf{X}\upp\bSigma_{\btheta}^{-1}\mathbf{X})^{-1}\mathbf{X}\upp\bSigma_{\btheta}^{-1}\mathbf{w}$.  Then, both cases are free of $\bbeta$,
\begin{linenomath*}
$$
[\mathbf{w}|\bX,\bSigma_{\btheta}] = 
\frac{1}{C_n}\exp[(\mathbf{w} - \mathbf{X}\hat{\bbeta})\upp\bSigma_{\btheta}^{-1}(\mathbf{w} - \mathbf{X}\hat{\bbeta})],
$$
\end{linenomath*}
where for ML estimation $C_n = \sqrt{2\pi^{n/2}|\bSigma_{\btheta}|}$ and for REML estimation  \\ $C_n = \sqrt{2\pi^{(n - p)/2}|\bSigma_{\btheta}||\mathbf{X}\upp\bSigma_{\btheta}^{-1}\mathbf{X}|}$. Note that \citet{bonat_practical_2016} only considered the marginal likelihood integrated over $\bw$, and did not consider the likelihood where $\bbeta$ was also integrated out (as in REML estimation) or back-substituted (as in ML estimation).

To obtain the marginal distribution of the data and covariance parameters we need the integral,
\begin{linenomath*}
$$
	[\mathbf{y}| \bphi, \bX,\bSigma_{\btheta}] = \int_\mathbf{w}  [\mathbf{y}|\bg^{-1}(\mathbf{w}),\bphi][\mathbf{w}|\bX,\bSigma_{\btheta}]d\mathbf{w}.
$$
\end{linenomath*}
Let us denote $\ell(\mathbf{w},\cdot) = \log([\mathbf{y}|\bg^{-1}(\mathbf{w}),\bphi][\mathbf{w}|\bX,\bSigma_{\btheta}])$, and consider $\int e^{\ell(\mathbf{w},\cdot)} d\mathbf{w}$. Let $\mathbf{v}$ be the gradient vector with $i$th element
\begin{linenomath*}
$$
v_{i}(\bphi,\btheta) = \frac{\partial \ell(\mathbf{w},\cdot)}{\partial w_i},
$$
\end{linenomath*}
and let $\mathbf{H}$ be the Hessian matrix with $i,j$th element,
\begin{linenomath*}
$$
H_{i,j}(\bphi,\btheta) = \frac{\partial^2 \ell(\mathbf{w},\cdot)}{\partial w_i\partial w_j},
$$
\end{linenomath*}
where in both $v_{i}(\bphi,\btheta)$ and $H_{i,j}(\bphi,\btheta)$ we show dependence on parameters $\bphi$ and $\btheta$. Henceforth, We assume that the log-likelihood is sufficiently well-behaved so that $\mathbf{H}$ is negative definite.  Using the multivariate Taylor series expansion of $\ell(\mathbf{w},\cdot)$ around some point $\mathbf{a}$,
\begin{linenomath*}
$$
\int_\mathbf{w} e^{\ell(\mathbf{w},\cdot)} d\mathbf{w} \approx \int_\mathbf{w} e^{\ell(\mathbf{a},\cdot) + \mathbf{v}\upp(\mathbf{w} -\mathbf{a}) + 1/2(\mathbf{w} - \mathbf{a})\upp\mathbf{H}(\mathbf{w} - \mathbf{a})} d\mathbf{w}.
$$
\end{linenomath*}
Now if $\mathbf{a}$ is a value at which $\mathbf{v} = \mathbf{0}$, then
\begin{linenomath*}
$$
\int_\mathbf{w} e^{\ell(\mathbf{w},\cdot)} d\mathbf{w} \approx e^{\ell(\mathbf{a},\cdot)} \int_\bw e^{ -  
	1/2(\mathbf{w} - \mathbf{a})\upp(-\mathbf{H})(\mathbf{w} - \mathbf{a})} d\mathbf{w} = e^{\ell(\mathbf{a},\cdot)} (2\pi)^{n/2}|-\mathbf{H}_\mathbf{a}(\bphi,\btheta)|^{-1/2},
$$
\end{linenomath*}
where $\mathbf{H}_\mathbf{a}(\bphi,\btheta)$ indicates $\mathbf{H}$ evaluated at $\mathbf{a}$ and we again show its dependence on $\bphi$ and $\btheta$. The result on the most right-hand side is familiar from the normalizing constant of a multivariate Gaussian distribution.  Hence,
\begin{linenomath*}
\begin{equation} \label{eq:covparmmarg}
[\mathbf{y}|\bphi,\bX,\bSigma_{\btheta}] = \int_\mathbf{w} e^{\ell(\mathbf{w},\cdot)} d\mathbf{w} \approx  [\mathbf{y}|\bg^{-1}(\mathbf{a}),\bphi][\mathbf{a}|\bX,\bSigma_{\btheta}](2\pi)^{n/2}|-\mathbf{H}_\mathbf{a}(\bphi,\btheta)|^{-1/2}.
\end{equation}
\end{linenomath*}


\subsection{Marginal Maximum Likelihood for Covariance Parameters}

From \eqref{eq:covparmmarg} an approximate marginal maximum likelihood estimator for $\bphi$ and $\btheta$ depending on finding $\mathbf{a}$ is
\begin{linenomath*}
\begin{equation} \label{eq:m2LLmargMLE}
\{\hat{\bphi}, \hat{\btheta} \} = \underset{\bphi,\btheta}{\arg\max} \left( \log[\mathbf{y}|\bg^{-1}(\mathbf{a}),\bphi] +
	\log[\mathbf{a}|\bX,\bSigma_{\btheta}] - \frac{1}{2}\log \left(|-\mathbf{H}_\mathbf{a}(\bphi,\btheta)|\right) \right),
\end{equation}
\end{linenomath*}
where we drop terms that do not contain $\bphi$ or $\btheta$.  Note that $\log[\mathbf{a}|\bX,\bSigma_{\btheta}]$ has the same form of the log-likelihood for ML or REML as in standard Gaussian models, but here it is evaluated at $\mathbf{a}$, where for ML
\begin{linenomath*}
\begin{equation} \label{eq:MLEa}
\log[\mathbf{a}|\bX,\bSigma_{\btheta}] = -\frac{n}{2}\log(2\pi) - \frac{1}{2}\log|\bSigma_{\btheta}| - \frac{1}{2}(\mathbf{a} - \bX\hat{\bbeta}_{\ba})\upp\bSigma_{\btheta}\upi(\mathbf{a} - \bX\hat{\bbeta}_{\ba}),
\end{equation}
\end{linenomath*}
and for REML
\begin{linenomath*}
\begin{equation} \label{eq:REMLa}
\log[\mathbf{a}|\bX,\bSigma_{\btheta}] = -\frac{n-p}{2}\log(2\pi) - \frac{1}{2}\log|\bSigma_{\btheta}| - \frac{1}{2}\log|\bX\upp\bSigma_{\btheta}^{-1}\bX| - \frac{1}{2}(\mathbf{a} - \bX\hat{\bbeta}_{\ba})\upp\bSigma_{\btheta}\upi(\mathbf{a} - \bX\hat{\bbeta}_{\ba}),
\end{equation}
\end{linenomath*}
where in both cases $\hat{\bbeta}_{\ba} = (\mathbf{X}\upp\bSigma_{\btheta}^{-1}\mathbf{X})^{-1}\mathbf{X}\upp\bSigma_{\btheta}^{-1}\mathbf{a}$. The result \eqref{eq:m2LLmargMLE} depends on finding $\mathbf{a}$ such that $\mathbf{v} = \mathbf{0}$.  To achieve this, we use Newton-Raphson, conditional on $\bphi$ and $\btheta$, which we describe next.

Assuming the conditional independence of the elements of $\mathbf{y}$ given $\bg^{-1}(\mathbf{w})$, we have
\begin{linenomath*}
\begin{equation} \label{eq:ll_plus_C}
\log([\mathbf{y}|\bg^{-1}(\mathbf{w}),\bphi][\mathbf{w}|\bX,\bSigma_{\btheta}]) = \sum_{i = 1}^n \log[y_i|g_i^{-1}(w_i),\bphi] - \frac{1}{2}(\mathbf{w} - \mathbf{X}\hat{\bbeta})\upp\bSigma_{\btheta}^{-1}(\mathbf{w} - \mathbf{X}\hat{\bbeta}) + C,
\end{equation}
\end{linenomath*}
where $C$ comprises terms that do not contain $\mathbf{w}$. Let $\mathbf{d}_{\bphi}$ be the vector with $i$th component,
\begin{linenomath*}
$$
d_{i} \equiv \frac{\partial\log[y_i|g_i^{-1}(w_i),\bphi]}{\partial w_i},
$$
\end{linenomath*}
and note that
\begin{linenomath*}
$$
\frac{\partial [-\frac{1}{2}(\mathbf{w} - \mathbf{X}\hat{\bbeta})\upp\bSigma_{\btheta}^{-1}(\mathbf{w} - \mathbf{X}\hat{\bbeta})]}{\partial \mathbf{w}} = -\bSigma_{\btheta}^{-1}\mathbf{w} + \bSigma_{\btheta}^{-1}\bX\hat{\bbeta},
$$
\end{linenomath*}
so the gradient of \eqref{eq:ll_plus_C} is
\begin{linenomath*}
\begin{equation} \label{eq:gdef}
\mathbf{v} = \mathbf{d}_{\bphi} - \bSigma_{\btheta}^{-1}\mathbf{w} + \bSigma_{\btheta}^{-1}\mathbf{X}\hat{\bbeta} = \mathbf{d}_{\bphi} - \mathbf{P}_{\btheta}\mathbf{w},
\end{equation}
\end{linenomath*}
where $\mathbf{P}_{\btheta} = \bSigma_{\btheta}^{-1} - \bSigma_{\btheta}^{-1}\mathbf{X}(\mathbf{X}\upp\bSigma_{\btheta}^{-1}\mathbf{X})^{-1}\mathbf{X}\upp\bSigma_{\btheta}^{-1}$. For the Hessian, let $\mathbf{D}_{\bphi}$ be a diagonal matrix with $i$th diagonal element,
\begin{linenomath*}
\begin{equation} \label{eq:Dii}
D_{i,i} \equiv \frac{\partial^2\log[y_i|g_i^{-1}(w_i),\bphi]}{\partial w_i^2},
\end{equation}
\end{linenomath*}
where all off-diagonal elements are zero because all second-order partial derivatives are 0 when $i \neq j$ due to conditional independence.  Then the Hessian of \eqref{eq:ll_plus_C} is
\begin{linenomath*}
\begin{equation} \label{eq:Hdef}
\mathbf{H} = \mathbf{D}_{\bphi} -\mathbf{P}_{\btheta}.
\end{equation}
\end{linenomath*}
Note that \eqref{eq:gdef} and \eqref{eq:Hdef} differ from the gradient and Hessian in \citet{ bonat_practical_2016} because we used $\hat{\bbeta}$ in \eqref{eq:ll_plus_C}, which contains $\bw$, whereas they used $\bbeta$.  In fact, the gradient and Hessian in \citet{ bonat_practical_2016} may be obtained from \eqref{eq:gdef} and \eqref{eq:Hdef} by replacing $\bP_{\btheta}$ with $\bSigma_{\btheta}^{-1}$.  This is an important difference, as it allows us to optimize for just the covariance parameters without having to do so for $\bbeta$ simultaneously.

A table of ${d}_{i}$ and ${D}_{i,i}$ for a few common distributions and link functions is given in Table~\ref{tab:dDlinkDist}.  In Table~\ref{tab:dDlinkDist}, we use alternative parameterizations for the negative binomial, gamma, and beta distributions so that $\textrm{E}(y) = \mu$.  We also reparameterize the inverse Gaussian distribution, and details for all distributions are given in the Appendix.

\begin{table}[H] 
	\caption{Flexibility of the HGLMM, showing how different distributions can be matched with different patterned covariance matrices. We also show distributions, inverse link functions, and first and second partial derivatives with respect to $w_{i}$ for various parts of the log-likelihood.  \label{tab:dDlinkDist}}
\begin{center}
\begin{tabular}{|cccc||c|}
\hline
\hline
\multicolumn{2}{|c}{$\log[\by|\bg^{-1}(\bw),\bphi]$} & \multicolumn{2}{c||}{$-(1/2)\log|-\bH_{\ba}(\bphi,\btheta)|$} & $+\log[\mathbf{a}|\bX,\bSigma_{\btheta}]$ \\
\hline
Distribution & $\bmu = \bg^{-1}(\bw)$ & ${d}_{i}$ & ${D}_{i,i}$ & $\bSigma_{\btheta}$-types\\
\hline
Binomial & $\bmu =\frac{\exp(\bw)}{1+\exp(\bw)}$ & $y_{i} - \frac{n_{i}\exp(w_{i})}{1+\exp(w_{i})}$ & $- \frac{n_{i}\exp(w_{i})}{(1+\exp(w_{i}))^{2}}$ & Random Effects \\ 
Poisson & $\bmu = \exp(\bw)$ & $y_{i} - \exp(w_{i})$ & $- \exp(w_{i})$ & Geostatistical \\ 
Neg. Binomial & $\bmu = \exp(\bw)$ &  $
\frac{\phi(y_{i} - e^{w_{i}})}{\phi + e^{w_{i}}}$ & $-\frac{\phi e^{w_{i}}(\phi + y_{i})}{(\phi + e^{w_{i}})^{2}}$ & Spatial Areal \\ 
Gamma & $\bmu = \exp(\bw)$ & $ -\phi + y_i \phi e^{-w_i}$ & $- y_i \phi e^{-w_i}$ & Time Series \\
Inv. Gaussian & $\bmu = \exp(\bw)$ & $\phi\left(\frac{y}{2e^{w_{i}}} - \frac{ e^{w_{i}}}{2y}\right) + \frac{1}{2}$ & $-\frac{\phi(e^{2w_{i}} + y_{i}^{2})}{2 y e^{w_{i}} }$ & Spatio-temporal\\
Beta & $\bmu =\frac{\exp(\bw)}{1+\exp(\bw)}$ & $\frac{-\phi e^{w_{i}}k_{0}(w_{i}|\phi,y_{i})}{(e^{w_{i}} + 1)^{2}}$ & $\frac{-\phi e^{2w_{i}}k_{1}(w_{i}|\phi,y_{i})}{(e^{w_{i}} + 1)^{4}}$ & {}\\
\hline
\hline
\end{tabular}
\end{center}
\vspace{-.7cm}
\begin{singlespace}
{\footnotesize
\-\hspace{.7cm} $k_{0}(w_{i}|\phi,y_{i}) = \psi^{(0)}\left(\frac{\phi e^{w_{i}}}{1 + e^{w_{i}}}\right) - \psi^{(0)}\left(\frac{\phi }{1 + e^{w_{i}}}\right)  + \log\left(\frac{1}{y_{i}} - 1\right)$ \\
\-\hspace{.7cm} $k_{1}(w_{i}|\phi,y_{i}) = \phi\left(\psi^{(1)}\left(\frac{\phi e^{w_{i}}}{1 + e^{w_{i}}}\right) + \psi^{(1)}\left(\frac{\phi }{1 + e^{w_{i}}}\right)\right) - 2\sinh(w_{i})\left(k_{0}(w_{i}|\phi,y_{i}) + 2\tanh\upi(1 - 2y_{i})\right)$ \\
\-\hspace{.7cm} $\psi^{(n)}(\cdot)$ is the $n$th derivative of the digamma function \\
\-\hspace{.7cm} $\sinh$ and $\tanh$ are the hyperbolic sine and tangent functions, respectively }
\end{singlespace}
\end{table}
%
%

Conditional on $\bphi$ and $\btheta$, a Newton-Raphson update is,
\begin{linenomath*}
\[
\mathbf{w}^{[k+1]} = \mathbf{w}^{[k]} - \mathbf{H}^{-1}\mathbf{v},
\]
\end{linenomath*}
and upon convergence we set $\mathbf{a} = \mathbf{w}$ in \eqref{eq:m2LLmargMLE} for any evaluation of the likelihood for given $\bphi$ and $\btheta$.  Notice that this makes the marginal maximum likelihood doubly iterative, as we solve for $\mathbf{a}$ while optimizing for $\bphi$ and $\btheta$. It is possible to use other maximization routines, such as the EM algorithm, but generally the Newton-Raphson algorithm converges rapidly (often around 10 iterations in our experience), and this was favored by \citet{ bonat_practical_2016} also.  However, on occasion, the stepsize needs to be adjusted so that $\mathbf{v}$ does not diverge.  For example, it is easy and fast to check $\mathbf{v}^{[k + 1]} = \mathbf{d}_{\bphi} - \mathbf{P}_{\btheta}\mathbf{w}^{[k+1]}$, and if $\mathbf{v}^{[k + 1]}$ is ``larger'' than $\mathbf{v}$ by some criterion (e.g., largest or average element of $\mathbf{v}$), then take
\begin{linenomath*}
\[
\mathbf{w}^{[k+1]} = \mathbf{w}^{[k]} - \alpha\mathbf{H}^{-1}\mathbf{v},
\]
\end{linenomath*}
where $0 < \alpha < 1$.  In the simulations below, we check $\mathbf{v}^{[k + 1]}$ in the manner described above, and set $\alpha =0.1$ if the largest element of $\mathbf{v}^{[k + 1]}$ is larger than the largest element of $\mathbf{v}$.  The advantage of using Newton-Raphson is that it provides $\mathbf{H}$, which is useful for making adjustments to variances when estimating fixed effects and making predictions, which we describe in the next section.

In summary, estimation of covariance parameters and $\bw$ can be written in the following steps,
\begin{enumerate}
	\item Get initial values for covariance parameters $\bphi$ and $\btheta$.  For example, for variance components, such as $\sigma^{2}_{0}$ and $\sigma^{2}_{1}$, apportion $\var(g(\by))$ equally among each variance component.  If there are many explanatory variables, a linear model can be fit to $g(\by))$ and residual variance could be used.
	\item Pick initial values for $\bw$. For example, set $\bw = g(\by)$ or as the residuals from a linear model fit to $g(\by)$.
	\item Use Newton-Raphson to estimate $\bw = \ba$ for given $\bphi$ and $\btheta$ in \eqref{eq:m2LLmargMLE}.
	\item Evaluate the log-likelihood in \eqref{eq:m2LLmargMLE} for $\bw$, $\bphi$ and $\btheta$.
	\item Loop through steps 3 and 4 for different values of $\bphi$ and $\btheta$ while optimizing for the log-likelihood in step 4 until convergence.
	
\end{enumerate}


\subsection{Inference for Fixed Effects}

In order to estimate $\bphi$ and $\btheta$ it was necessary to optimize the likelihood for $\mathbf{w}$, which we called $\mathbf{a}$, using Newton-Raphson, for each evaluation of the likelihood.  Upon convergence in estimating $\bphi$ and $\btheta$, we also have optimized $\mathbf{w}$. Let us denote the optimizing value as $\hat{\mathbf{w}} = \mathbf{a}$.  

\citet{ bonat_practical_2016} proposed profile likelihood for estimating $\bbeta$ and obtaining confidence intervals, but their proposal is computationally demanding and does not extend well to cases with many coefficients in $\bbeta$. An alternative estimator of $\bbeta$ may be obtained by replacing the unobserved ${\bf w}$ with its predicted value $\hat{\bf w}={\bf a}$, obtained as described in the previous subsection, in the expression for what would be the generalized least squares estimator of $\bbeta$ if in fact ${\bf w}$ {\em was} observed.  This yields the estimator
$\hat{\bbeta} = \mathbf{B}\hat{\mathbf{w}}$, where $\mathbf{B} = (\mathbf{X}\upp\bSigma_{\btheta}^{-1}\mathbf{X})^{-1}\mathbf{X}\upp\bSigma_{\btheta}^{-1}$.  In order to estimate the variance of $\hat{\bbeta}$, it is convenient to condition on $\mathbf{w}$ as if it was observed, and then use the following well-known result, often called the law of total variance,
\begin{linenomath*}
\begin{equation} \label{eq:evarvare}
\var(\hbw) = \textrm{E}_{\bw}[\var(\hbw|\bw)] + \var_{\bw}[\textrm{E}(\hbw|\bw)].
\end{equation}
\end{linenomath*}
Due to the optimization of $\hbw$ from the likelihood, we will assume that $\hbw|
\bw$ is approximately distributed as $\textrm{N}(\bw,\bF_{\bw}\upi$), where $\bF_{\bw}$ is the observed Fisher information, or, less strictly, that $\textrm{E}(\hbw|\bw) = \bw$ and $\var(\hbw|\bw) = \bF_{\bw}\upi$, approximately. Thus, for the second term in \eqref{eq:evarvare}, $\var_{\mathbf{w}}[\textrm{E}(\hat{\bw}|\bw)] = \bSigma_{\btheta}$, which we approximate by $\bSigma_{\hbtheta}$ after substituting estimated parameters $\hbtheta$ for $\btheta$. For the first term in \eqref{eq:evarvare}, the observed Fisher information is equivalent to $-\mathbf{H}_{\bw}(\bphi,\btheta)^{-1}$, where we show the dependence on parameters $\bphi$ and $\btheta$, and on $\bw$ that comes from $\bD_{\bphi}$ in \eqref{eq:Hdef} (see the examples of $D_{i,i}$ in Table~\ref{tab:dDlinkDist}). To obtain the Fisher information would require taking the expectation, $\textrm{E}_{\bw}[\var(\hbw|\bw)]$, but this is complicated, and \citet{efron_assessing_1978} argue for using the observed Fisher information instead, so we simply replace $\bw$ in $-\mathbf{H}_{\bw}(\bphi,\btheta)^{-1}$ with $\hat{\mathbf{w}} = \mathbf{a}$, and we also replace $\bphi$ and $\btheta$ by their estimates $\hat{\bphi}$ and $\hbtheta$, and denote this as $-\mathbf{H}_{\hbw}(\hat{\bphi},\hat{\btheta})^{-1}$. Then an estimator of the covariance matrix of estimated fixed effects is
\begin{linenomath*}
\begin{equation} \label{eq:sglmm_varfe}
\widehat{\var}(\hat{\bbeta}) = \bB [\var(\hbw)] \bB\upp = \mathbf{B}[-\mathbf{H}_{\hat{\bw}}(\hat{\bphi},\hat{\btheta})^{-1}]\mathbf{B}\upp + \mathbf{C}_{\hat{\bbeta}},
\end{equation}
\end{linenomath*}
where $\mathbf{C}_{\hat{\bbeta}} = \mathbf{B}\bSigma_{\hbtheta}\mathbf{B}\upp$, which simplifies to $\mathbf{C}_{\hbbeta} = (\mathbf{X}\upp\bSigma_{\hbtheta}^{-1}\mathbf{X})^{-1}$, the usual estimated variance-covariance matrix of fixed effects when using generalized least squares if $\bw$ were observed.


\subsection{Inference for Prediction} \label{sec:pred_infer}

So far, we have estimated $\btheta$, $\bphi$, and $\bbeta$, predicted $\bw$, and obtained estimated covariance matrices for $\hat{\bbeta}$ and $\hat{\bw}$.  Now let us consider the task of prediction for unsampled data, which may be in space, or time, or by design.  We will denote the unsampled $\{w_{i}\}$ by the vector $\bu$.  We can extend the linear model \eqref{eq:lm_eta} as
\begin{linenomath*}
\begin{equation} \label{eq:lm_4pred}
	\left( \begin{array}{c}
		\bw \\
		\bu{}
	\end{array} \right) \sim 
	N\left(\left( \begin{array}{c}
		\bX\bbeta \\
		\bX_{\bu}\bbeta
	\end{array} \right), 
		\left( \begin{array}{cc}
		\bSigma_{\btheta} & \bSigma_{\bf wu} \\
		\bSigma\upp_{\bf wu} & \bSigma_{\bf uu}
	\end{array} \right)\right).   
\end{equation}
\end{linenomath*}
Our goal is the prediction of $\bu$.   If ${\bf w}$ was observed, the best linear unbiased predictor (BLUP) of $\bu$ would be $\bLambda\mathbf{w}$, where $\bLambda = \mathbf{X}_{\mathbf{u}}\mathbf{B} + \bSigma\upp_{\mathbf{w}\mathbf{u}}\bSigma^{-1}_{\btheta} - \bSigma\upp_{\mathbf{w}\mathbf{u}}\bSigma^{-1}_{\btheta}\mathbf{X}\mathbf{B}$ \citep{goldberger_best_1962}.  Since, however, ${\bf w}$ is unobserved, an alternative predictor of ${\bf u}$ may be obtained by substituting $\hat{\bf w}$ for ${\bf w}$ in this expression, yielding $\hat{\bu}=\bLambda\hat{\bf w}$.  

To determine the sampling properties of $\hat{\bf u}$, we again we need to make some adjustments that account for the substitution of $\hat{\mathbf{w}}$ for $\mathbf{w}$ in the BLUP (and for now, we are conditioning on covariance parameters when creating the covariance matrices).  Assuming again that $\hat{\bf w}$ is unbiased for ${\bf w}$, it is easily seen that this alternative predictor is unbiased, i.e., E$(\bLambda\hat{\bf w})={\rm E}({\bf u})$.  Now we want an estimator of the prediction error variance associated with this predictor, which is $\var(\hat{\mathbf{u}} -\mathbf{u}) = \var(\bLambda\hat{\mathbf{w}} - \mathbf{u})$.  Note that the prediction error variance of the BLUP is
\begin{linenomath*}
\begin{equation} \label{eq:UK_MSPE}
\var(\bLambda{\bw} - \bu) = \bSigma_{\bu\bu} - \bSigma_{\bw\bu}\upp\bSigma_{\btheta}\upi\bSigma_{\bw\bu} + \bK\bC_{\bbeta}\bK\upp,
\end{equation}
\end{linenomath*}
where $\mathbf{K} = \mathbf{X}_{\mathbf{u}} - \bSigma\upp_{\mathbf{w}\mathbf{u}}\bSigma^{-1}_{\btheta}\mathbf{X}$ \citep{goldberger_best_1962}. To obtain the prediction error variance of our alternative predictor, it is convenient to condition on $\mathbf{w}$ and $\mathbf{u}$ as we did in \eqref{eq:evarvare}, i.e.,
\begin{linenomath*}
$$
\var(\bLambda\hat{\mathbf{w}}-\mathbf{u}) = \textrm{E}_{\mathbf{w},\mathbf{u}}[\var(\bLambda\hat{\mathbf{w}} -\mathbf{u}|\mathbf{w},\mathbf{u})] + \var_{\mathbf{w},\mathbf{u}}[\textrm{E}(\bLambda\hat{\mathbf{w}}-\mathbf{u}|\mathbf{w},\mathbf{u})].
$$
\end{linenomath*}
Owing to the assumed unbiasedness of $\hat{\mathbf{w}}$ for $\mathbf{w}$, we have  $\textrm{E}(\bLambda\hat{\mathbf{w}}-\mathbf{u}|\mathbf{w},\mathbf{u}) = \bLambda\mathbf{w}-\mathbf{u}$, and the variance of this is given by \eqref{eq:UK_MSPE}. Conditionally, $\var_{\hat{\mathbf{w}}}(\bLambda\hat{\mathbf{w}}-\mathbf{u})$ does not depend on $\mathbf{u}$, so $\textrm{E}_{\mathbf{w},\mathbf{u}}[\var(\bLambda\hat{\mathbf{w}} -\mathbf{u}|\mathbf{w},\mathbf{u})] = \textrm{E}_{\mathbf{w}}(\bLambda[-\mathbf{H}_{\bw}(\bphi,\btheta)^{-1}]\bLambda\upp)$, and, as we did in the previous section, rather than take expectation, we simply use the observed Fisher information by replacing $\mathbf{w}$ in $\mathbf{H}$ with its estimator $\hat{\bw} = \mathbf{a}$ and replacing $\bphi$ and $\btheta$ with $\hat{\bphi}$ and $\hat{\btheta}$.  Putting them together, we obtain
\begin{linenomath*}
\begin{equation} \label{eq:sglmm_varpred}
\widehat{\var}(\bLambda\hat{\mathbf{w}}-\mathbf{u}) = \bLambda[-\mathbf{H}_{\hat{\bw}}(\hat{\bphi},\hat{\btheta})^{-1}]\bLambda\upp + \bSigma_{\mathbf{u}\mathbf{u}} - \bSigma_{\mathbf{w}\mathbf{u}}\upp\bSigma_{\hat{\btheta}}^{-1}\bSigma_{\mathbf{w}\mathbf{u}} + \mathbf{K}\mathbf{C}_{\hat{\bbeta}}\mathbf{K}\upp.
\end{equation}
\end{linenomath*}
All covariance matrices depend on $\btheta$, although the notation makes this explicit only for the covariance matrix of ${\bf w}$.  We replace $\btheta$ in these matrices by its estimator $\hat{\btheta}$, where the fitted covariance function that is used to estimate $\bSigma_{\btheta}$ is also used to estimate $\bSigma_{\bw\bu}$ and $\bSigma_{\bu\bu}$.


\section{Simulations}

\begin{figure}[H]
  \begin{center}
	    \includegraphics[width=.95\linewidth]{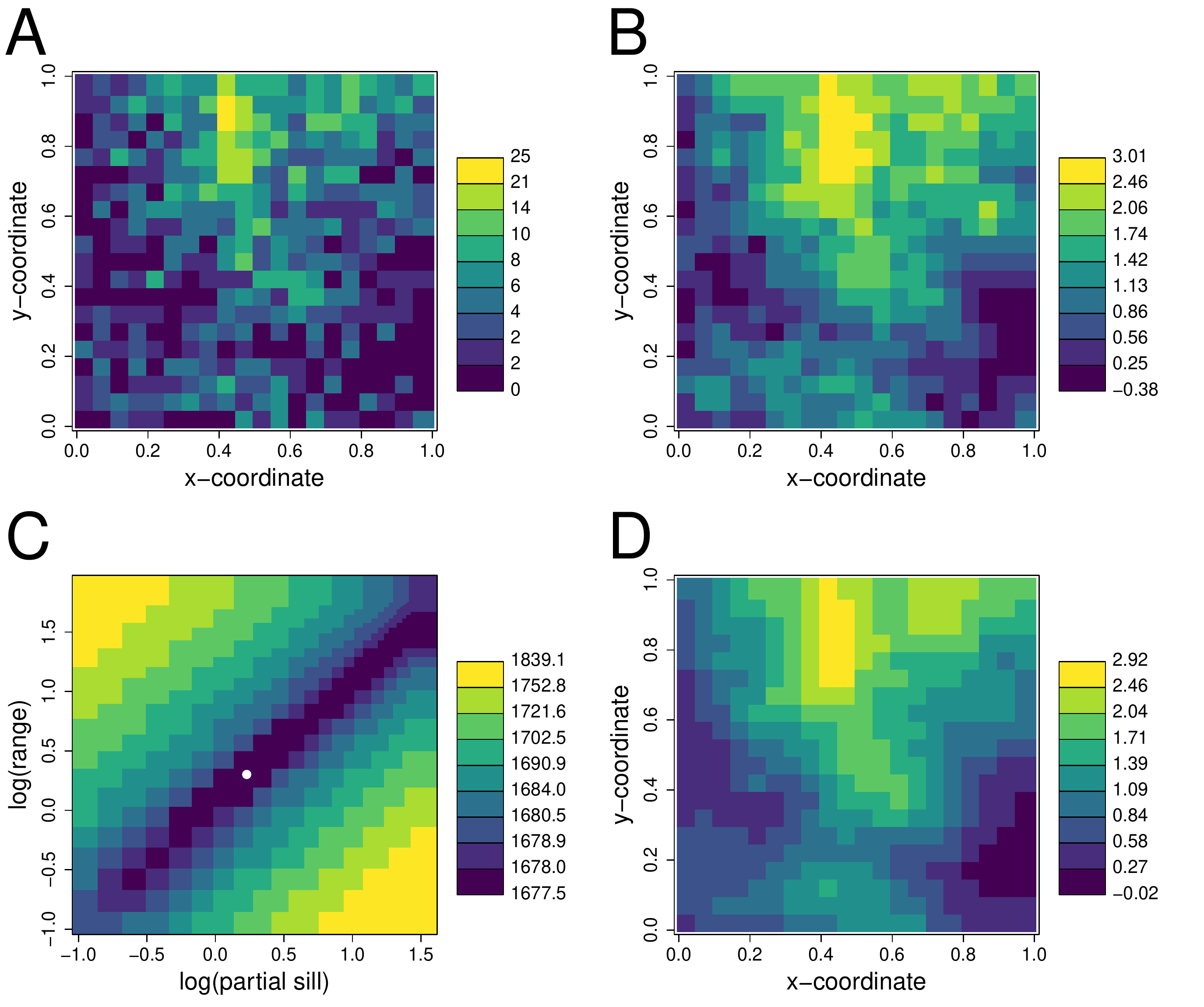}
  \end{center}
  \caption{Estimation for simulated Poisson count data. A. Simulated count data using the model described in the text.  B. The true simulated $\mathbf{w}$ values. C. The marginal likelihood surface of the simulated data.  The white circle shows the estimated value.  D. The estimated $\hat{\mathbf{w}}$ values. \label{Fig:sglm_likelihood_estimation}}
\end{figure}

We first illustrate our methods with a simulated spatial dataset so that we know the true values of all parameters and ${\bf w}$.  We created a square grid of $20 \times 20$ data locations on a $(0,1) \times (0,1)$ (unit square) domain. On this grid, we generated a single $\mathbf{w}$ from a multivariate normal distribution with mean vector ${\bf 1}\bbeta_0$, where $\beta_0=2$, and covariance matrix determined by evaluating an exponential autocovariance model, $\cov(w(\bs_{i}),w(\bs_{j})) = \sigma^{2}_{1}\exp(-\delta_{i,j}/\rho) + \sigma^{2}_{0}\mathcal{I}(\delta_{i,j} = 0)$, where $\bs_{i}$ $(i=1,\ldots,400)$ is the vector of spatial coordinates at location $i$, $\delta_{i,j}$ is Euclidean distance between the $i$th and $j$th locations,  and $\mathcal{I}(\cdot)$ is the indicator function, equal to one if its argument is true and equal to 0 otherwise.  We set $\sigma^{2}_1 = 1$, $\rho = 1$, and $\sigma^{2}_{0} = 0.0001$.  The 400 simulated values in $\mathbf{w}$ are shown in Figure~\ref{Fig:sglm_likelihood_estimation}B.  Conditional on $\mathbf{w}$, at each data location we independently simulated a Poisson random variable with mean equal to $\exp(w_{i})$, and these are shown in Figure~\ref{Fig:sglm_likelihood_estimation}A.

Using the simulated data in Figure~\ref{Fig:sglm_likelihood_estimation}A and assuming that the parameters of the mean and exponential covariance function were unknown, we optimized the likelihood in \eqref{eq:m2LLmargMLE} for $\btheta = (\sigma_{1}^{2},\rho, \sigma_{0}^{2})$ using REML.  The parameter estimates so obtained were $\hat{\sigma}_{1}^{2} = 1.247$, $\hat{\rho} = 1.341$, and $\hat{\sigma}_{0}^{2} = 1.392 \times 10^{-11}$.  The likelihood surface for $\sigma_{1}^{2}$ and $\rho$ is shown in Figure~\ref{Fig:sglm_likelihood_estimation}C. A pronounced ridge reveals a positive association in the likelihood between $\sigma_{1}^{2}$ and $\rho$, which is typical for geostatistical models. The estimation of $\btheta = (\sigma_{1}^{2},\rho, \sigma_{0}^{2})$ also provided $\hat{\mathbf{w}}$, which is shown in Figure~\ref{Fig:sglm_likelihood_estimation}D, and it appears that we were able to recover the spatial patterning within the true simulated $\mathbf{w}$ quite well.  The estimated value of the overall intercept, $\beta_0$, was 0.852 with an estimated standard error of 0.865.

\begin{spacing}{0.8}
\vspace{.4cm}
\noindent --------------------------------------------- \\
\textbf{Figure~\ref{Fig:sglm_likelihood_estimation} here}: Estimation for simulated Poisson count data. A. Simulated count data using the model described in the text.  B. The true simulated $\mathbf{w}$ values. C. The marginal likelihood surface of the simulated data.  The white circle shows the estimated value.  D. The predicted $\hat{\mathbf{w}}$ values. \\
--------------------------------------------- \\ 
\end{spacing}

Of course, this is just one simulation. In order to evaluate the bias of the estimators and the coverage of confidence intervals, we conducted a larger simulation experiment. We simulated data at 200 data locations chosen randomly within the unit square. We simulated from a multivariate normal distribution with exactly the same autocovariance model used before, but with mean structure
\begin{linenomath*}
$$
\textrm{E}(w_{i}) = \beta_{0} + \beta_{1}x_{i} + \beta_{2}\tau_{i} + \beta_{3}x_i\tau_i
$$
\end{linenomath*}
where $x_{i}$ was randomly and independently simulated from $\textrm{N}(0,1)$, $\tau_{i}$ was randomly and independently simulated as a Bernoulli variable with probability $p = 0.5$
We set $\bbeta = (0.5, 0.5, -0.5, 0.5)\upp$.  We also created 100 prediction locations on a $10 \times 10$ square grid within the unit square, and explanatory variables were also simulated at the prediction locations. In total, 300 $w_{i}$ values (200 observed, 100 for prediction) were simulated from a model as given in \eqref{eq:lm_4pred}. We then created the observed data as counts from a Poisson distribution conditional on $\mathbf{w}$, where at each of the 200 data locations we independently simulated a Poisson random variable with mean equal to $\exp(w_{i})$.  In this manner, we simulated 2000 data sets of size 300 to assess bias and confidence/prediction interval coverage.

For each simulated data set, we first estimated the covariance parameters using \eqref{eq:m2LLmargMLE}, where $\log[\mathbf{a}|\bX,\bSigma_{\btheta}]$ is the REML log-likelihood given by \eqref{eq:REMLa}, and again we used an exponential autocovariance model for fitting. We set upper bounds on the parameter spaces for $\sigma^{2}_{1}$ and $\rho$, at 10 times $\var(g(y))$ and 10 times the maximum distance among all spatial locations, respectively.  We did this because sometimes either the estimate of $\sigma^{2}_1$ or $\rho$, or both, appears to increase without bound, corresponding to a very flat ridge in the likelihood similar to that seen in Figure~\ref{Fig:sglm_likelihood_estimation}C, and only their ratio is important for estimation and prediction \citep{zhang_inconsistent_2004}. This stabilized estimation over so many simulations.

We used the estimated covariance parameters as plug-in values for the autocovariance model to obtain $\bSigma_{\hat{\btheta}}$, and, along with the estimated $\hat{\mathbf{w}}$, we estimated fixed effects by $\hat{\bbeta} = \mathbf{B}\hat{\mathbf{w}}$.  To estimate bias, we took the average of $\hat{\bbeta} - \bbeta$ (element-wise) over all 2000 simulated data sets, and to estimate mean-squared error (MSE) we took the average of $(\hat{\bbeta} - \bbeta)^2$ (element-wise) over all 2000 simulated data sets.  We also formed nominal 90\% confidence intervals as $\hat{\beta}_j \pm 1.645 \widehat{\textrm{se}}(\hat{\beta}_j)$, where $\widehat{\textrm{se}}(\hat{\beta}_j)$ was the square root of the $(j+1)$th diagonal element of \eqref{eq:sglmm_varfe} ($j=0,1,2,3$).  We computed the proportion of times, over the 2000 simulations that the confidence intervals contained the true values.  If we are estimating the $\beta_j$s and their variances well, the empirical coverage should be close to 90\%.  We also computed the confidence interval coverage based on using the diagonal elements of $\mathbf{C}_{\hat\beta}$, the unadjusted GLS variance estimator.  

The results are shown in Table~\ref{tab:sglm_fe}, where we see that there is very little bias in estimating any of the parameters in $\bbeta$.  Recalling that MSE is the sum of squared bias and variance, it is clear that bias was a very small component of MSE in these simulations.  The confidence interval coverage for $\beta_{0}$ is low, but estimating the overall intercept is difficult for normal spatial models as well.  The confidence interval coverages for $\beta_{1}$, $\beta_2$, and $\beta_{3}$ are very close to 90\% when using \eqref{eq:sglmm_varfe}, but they are substantially less than 90\% when using only $\mathbf{C}_{\hat{\bbeta}}$.

We also used the estimated covariance parameters in $\bSigma_{\hat{\btheta}}$ and the predicted $\hat{\mathbf{w}}$ to make predictions, using $\hat{\mathbf{u}} = \bLambda\hat{\mathbf{w}}$, at all 100 prediction locations for each simulated data set.  To estimate prediction bias, we used the average of the elements of $\hat{\mathbf{u}} - \mathbf{u}$ for each simulated data set, where $\mathbf{u}$ contains the simulated values at the 100 prediction locations, and then averaged those across the 2000 simulated data sets.  We also formed 90\% prediction intervals as $\hat{u}_k \pm 1.645 \widehat{\textrm{se}}(\hat{u}_k)$ $(k=1,\ldots,100)$, where $\widehat{\textrm{se}}(\hat{u}_k)$ was the square root of the $k$th diagonal element of \eqref{eq:sglmm_varpred}.  We computed the proportion of times, over the 100 predictions and 2000 simulations, that the prediction intervals contained the true values, which should be about 90\%. Table~\ref{tab:sglm_fe} gives the prediction results, which show little indication of bias.  Coverage was very close to 90\% when using \eqref{eq:sglmm_varpred}, but too low when using the naive \eqref{eq:UK_MSPE}.

\begin{table}[H] 
	\caption{Bias and coverage for estimation of fixed effects $\bbeta$ and for prediction of $\mathbf{u}$ at unobserved locations.  MSE is mean-squared (prediction) error, and ratio is bias$^2$/MSE. Coverage is for 90\% confidence and prediction intervals, and CI90$_{c}$ used the corrected versions in \eqref{eq:sglmm_varfe} and \eqref{eq:sglmm_varpred}, while CI90$_{u}$ shows coverage using the uncorrected standard-error estimator based on $\mathbf{C}_{\bbeta}$ and the uncorrected prediction standard errors using \eqref{eq:UK_MSPE}. Values in bold are outside of the 99\% confidence interval for 2000 independent Bernoulli trials with true value 0.9.  \label{tab:sglm_fe}}
\begin{center}
\begin{tabular}{|c|rrrrr|}
\hline
\hline
effect & bias & MSE & ratio & CI90$_{u}$ & CI90$_{c}$ \\
\hline{}  
$\hbeta_{0}$ & 0.042 & 0.530 & 0.003 & \textbf{0.750} & \textbf{0.773} \\ 
$\hbeta_{1}$ & -0.005 & 0.008 & 0.003 & \textbf{0.367} & 0.899 \\ 
$\hbeta_{2}$ & 0.004 & 0.024 & 0.001 & \textbf{0.310} & 0.910 \\ 
$\hbeta_{3}$ & -0.005 & 0.018 & 0.001 & \textbf{0.357} & 0.911 \\ 
$\hat{\bu}$ & 0.039 & 0.138 & 0.011 & \textbf{0.700} & 0.898 \\ 
\hline
\hline
\end{tabular}
\end{center}
\end{table}

In addition to the Poisson distribution, we did similar simulations for all five of the other distributions in Table~\ref{tab:dDlinkDist}.  No $\phi$ was needed for the binomial (Bernoulli) distribution, but for the beta distribution we set $\phi = 10$, and for the negative binomial, gamma, and inverse Gaussian distributions we set $\phi = 1$. Our alternative estimators and predictors appeared to be unbiased in all five cases, so we only show the corrected 90\% confidence interval coverage in Table~\ref{tab:simCI90_all}.  In all but one case, the intervals have close to 90\% confidence and prediction interval coverage, the exception being the beta distribution, which undercovers slightly, especially for prediction.  More simulations will be necessary to fully characterize when the intervals corresponding to these distributions have shortcomings.

\begin{table}[H] 
	\caption{Interval coverage for estimation of fixed effects $\bbeta$ and for prediction of $\mathbf{u}$ at unobserved locations for the five distributions in Table~\ref{tab:dDlinkDist} that were not included in Table~\ref{tab:sglm_fe}; binomial (bino), beta, negative binomial (nbin), gamma (gamm) and inverse Gaussian (iGau).  Coverage is for 90\% confidence and prediction intervals, using the corrected versions in \eqref{eq:sglmm_varfe} and \eqref{eq:sglmm_varpred}.  Values in bold are outside of the 99\% confidence interval for 2000 independent Bernoulli trials with true value 0.9.  \label{tab:simCI90_all}}
\begin{center}
\begin{tabular}{|c|rrrrr|}
\hline
\hline
effect & bino & beta & nbin & gamm & iGua \\
\hline{}
$\hbeta_{0}$ & \textbf{0.712} & \textbf{0.923} & \textbf{0.730} & \textbf{0.839} & \textbf{0.708} \\ 
$\hbeta_{1}$ & 0.901 & \textbf{0.828} & 0.885 & 0.898 & 0.908 \\ 
$\hbeta_{2}$ & 0.914 & \textbf{0.837} & 0.898 & 0.894 & 0.888 \\ 
$\hbeta_{3}$ & 0.915 & \textbf{0.816} & 0.901 & 0.892 & 0.900 \\ 
$\hat{\bu}$ & 0.887 & \textbf{0.744} & 0.883 & 0.892 & \textbf{0.875} \\  
\hline
\hline
\end{tabular}
\end{center}
\end{table}


\section{Examples}

We demonstrate the methods with three example datasets that use all of the distributions in Table~\ref{tab:dDlinkDist}, combined with covariance matrices developed through spatial autoregressive models, time series models, geostatistical models, and variance components models that include random effects.


\subsection{1980 Presidential Turnout in Texas}

This dataset contains the proportion of the population over age 19 that cast votes in the 1980 presidential election in the United States.  The proportions are for each of the 254 counties in Texas.  The data for the whole of the United States were collected and reported in \citet{pace_quick_1997}, and are available in the \texttt{R} package \texttt{spData} \citep{bivand_roger_spdata_2023}.  We created a subset of the data for Texas only.  The response variable is reported as a proportion, but we also created a binary variable by assigning a value of 1 to those proportions greater than 0.5 and assigning a value of zero otherwise. Here, we will fit the binomial distribution (actually a Bernoulli distribution because all sample sizes are one) to the binary response variable, and the beta distribution to the proportional response variable.

\begin{figure}[H]
  \begin{center}
	    \includegraphics[width=.95\linewidth]{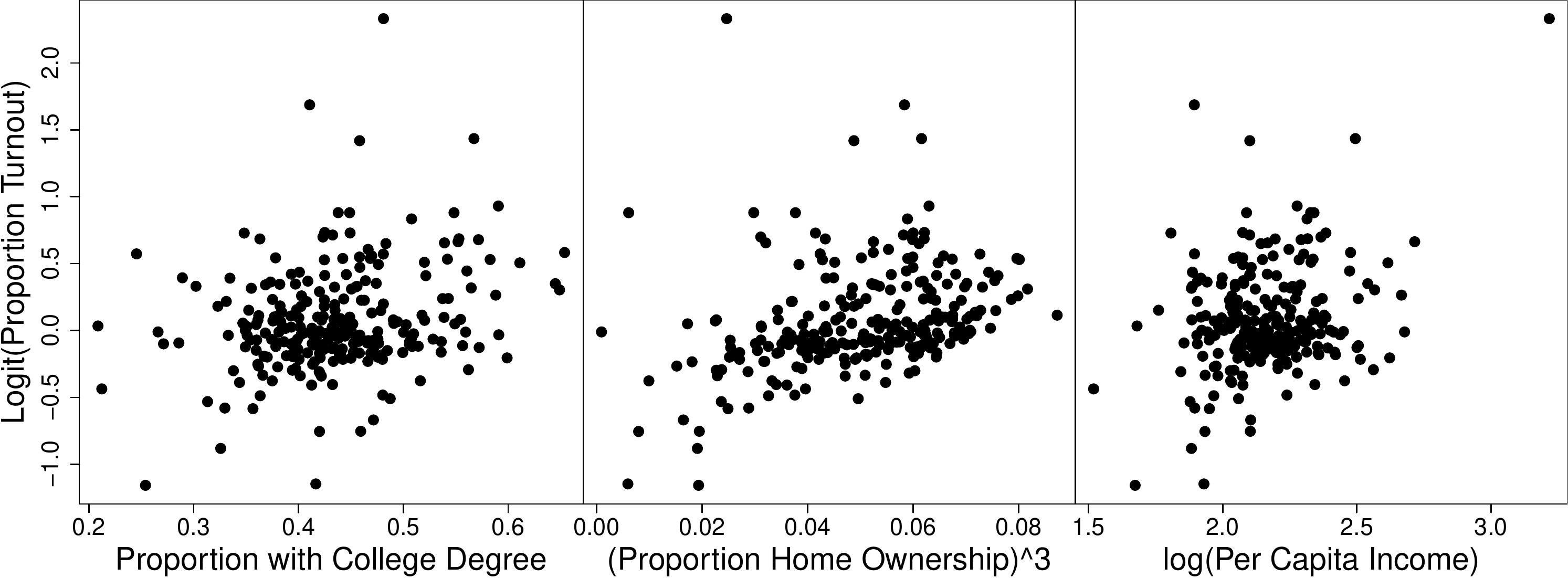}
  \end{center}
  \caption{Scatterplot of the logit of voter-turnout response variable by the three explanatory variables.  Note the transformations of some explanatory variables, where proportion of home ownership was cubed, and natural logs were taken of per capita income.  \label{Fig:TexTurn_scatter}}
\end{figure}

There are three explanatory variables in the data set: 1) proportion of population with college degrees, 2) proportion of home ownership, and 3) per capita income, where for all three variables the values are with respect to the total population over age 19 that were eligible to vote.  A scatterplot of the logit of the proportional response variable for all three explanatory variables is given in Figure~\ref{Fig:TexTurn_scatter}.  In an attempt to linearize relationships, we cubed the explanatory variable for the proportion of home ownership and took the natural logarithm of per capita income. The linear model that we consider then is,
\begin{linenomath*}
$$
\bw = \bX\bbeta + \br_{1},
$$
\end{linenomath*}
where $\bX$ contains a column of ones for an overall mean and three columns for the (transformed) explanatory variables.

\begin{spacing}{0.8}
\vspace{.4cm}
\noindent --------------------------------------------- \\
\textbf{Figure~\ref{Fig:TexTurn_scatter} here}: Scatterplot of the logit of voter-turnout response variable by the three explanatory variables.  Note the transformations of some explanatory variables, where proportion of home ownership was cubed, and natural logs were taken of per capita income. \\
--------------------------------------------- \\ 
\end{spacing}

For the spatial random effects $\br_{1}$, we fit two spatial autoregressive models to the data for the 254 counties, a conditional autoregressive (CAR) model and a simultaneous autoregressive (SAR) model.  These models rely on neighbor definitions, rather than distance directly.  We defined a neighbor of a county as any other county whose centroid was within 150 km of the originating county's centroid.  Using that definition, some counties had but a single neighbor, while others had many; the maximum was 38.  Let $\bW=(W_{i,j})$ denote a neighbor incidence matrix, where $W_{i,j}=1$ if county $j$ is a neighbor of county $i$, and $W_{i,j}=0$ otherwise, where the diagonal is all zeros (a site is not a neighbor of itself). Let $\bW_{rs}$ be a ``row-standardized'' version of $\bW$, where the elements in any row of $\bW$ are divided by their row sum, $W_{i,+} = \sum_{j}W_{i,j}$.  Then the covariance matrix for a CAR model is
\begin{linenomath*}
$$
\bSigma_{\btheta} = \sigma^{2}(\bI - \rho\bW_{rs})\upi\bM_{rs},
$$
\end{linenomath*}
where $\bM_{rs}$ is a diagonal matrix with $i$th diagonal element $1/W_{i,+}$.  The SAR covariance matrix is
\begin{linenomath*}
$$
\bSigma_{\btheta} = \sigma^{2}[(\bI - \rho\bW_{rs})(\bI - \rho\bW_{rs}\upp)]\upi.
$$
\end{linenomath*}
In both covariance matrices, $\rho$ is a spatial dependence parameter.

For the model with binary response variable, the three explanatory variables, and the CAR covariance matrix, using \eqref{eq:m2LLmargMLE} with REML we estimated $\sigma^{2} = 0.625$ and $\rho = 0.999$.  Changing the model's covariance matrix to the SAR covariance matrix, we estimated $\sigma^{2} = 0.265$ and $\rho = 0.964$.  The minimized value of the minus twice the log-likelihood in \eqref{eq:m2LLmargMLE} was 298.14 for the CAR model, while it was 279.15 for the SAR model, indicating that the SAR model was a better choice.   Table~\ref{tab:FE_bin} gives fixed effects estimates.  Note the large difference in standard errors using the naive approach based on $\bC_{\hbbeta}$ and the corrected version given by \eqref{eq:sglmm_varfe}.

\begin{table}[H] 
	\caption{Estimated fixed effects table for Texas turnout data using binary response variable.  The estimates are given by Est., while s.e.$_{u}$ is the naive standard error using only $\bC_{\bbeta}$ from Section 2.3 and s.e.$_{c}$ is the corrected standard error using \eqref{eq:sglmm_varfe}.  The t-val.\ is the estimate divided by the corrected standard error, and the p-val.\ is the probability of exceeding the t-value if the effect were truly zero.  \label{tab:FE_bin}}
\begin{center}
\begin{tabular}{|c|rrrrr|rr|}
\hline
\hline{}
{} & \multicolumn{5}{c|}{SAR model} & \multicolumn{2}{c|}{CAR model} \\
Effect & Est. & s.e.$_{u}$ & s.e.$_{c}$ & t-val. & p-val. & Est. & s.e.$_{c}$ \\
\hline{}
Intercept & -5.725 & 0.502 & 2.397 & 2.388 & 0.0177 & -2.898 & 1.975 \\ 
College & 4.439 & 0.701 & 3.694 & 1.202 & 0.2307 & 4.137 & 2.979 \\ 
Home-owner & 69.768 & 2.221 & 12.771 & 5.463 & $<$ 0.0001 & 56.635 & 10.318 \\ 
Income &  -0.171 & 0.260 & 1.414 & 0.121 & 0.9040 & -0.860 & 1.151\\ 
\hline
\hline
\end{tabular}
\end{center}
\end{table}

For the proportion turnout response variable, the beta distribution in Table~\ref{tab:dDlinkDist} was used, and for the CAR covariance matrix we estimated $\sigma^{2} = 0.312$, $\rho = 0.999$, and $\phi = 48.7$, while for the SAR covariance matrix we estimated $\sigma^{2} = 0.0126$, $\rho = 0.941$, and $\phi = 46.9$.  The minimized value of minus twice the log-likelihood in \eqref{eq:m2LLmargMLE} was -554.6 for the CAR model, while it was -557.8 for the SAR model, indicating the SAR model was a better choice, just as for the binary data. Table~\ref{tab:FE_turnout} gives fixed effects estimates.  As was the case for the binary data, there is a large difference in standard errors using the naive approach based on $\bC_{\hbbeta}$ and the corrected version given by \eqref{eq:sglmm_varfe}.  The overall patterns of coefficient estimates and their precisions are similar between SAR and CAR models in both Tables~\ref{tab:FE_bin} and \ref{tab:FE_turnout}.  In comparing Table~\ref{tab:FE_bin} to Table~\ref{tab:FE_turnout}, there appears to be more precision in the estimates in Table~\ref{tab:FE_turnout}, especially regarding the significance of the per capita income variable.  This is not surprising because the transformation of the proportional turnout data into binary data invariably results in a loss of information.

\begin{table}[H] 
	\caption{Estimated fixed effects table for Texas turnout data using proportional response variable.  The headings are the same as for Table~\ref{tab:FE_bin}.  \label{tab:FE_turnout}}
\begin{center}
\begin{tabular}{|c|rrrrr|rr|}
\hline
\hline{}
{} & \multicolumn{5}{c|}{SAR model} & \multicolumn{2}{c|}{CAR model} \\
Effect & Est. & s.e.$_{u}$ & s.e.$_{c}$ & t-val. & p-val. & Est. & s.e.$_{c}$ \\
\hline{}
Intercept & -1.614 & 0.105 & 0.253 & 6.379 & $<$0.0001 & -1.427 & 0.340\\ 
College & 0.407 & 0.154 & 0.407 & 1.000 & 0.3184 & 0.481 & 0.394\\ 
Home-owner & 8.711 & 0.486 & 1.300 & 6.703 & $<$0.0001 & 8.552 & 1.273\\ 
Income &  0.470 & 0.057 & 0.142 & 3.317 & 0.0010 & 0.390 & 0.143\\ 
\hline
\hline
\end{tabular}
\end{center}
\end{table}

\begin{figure}[H]
  \begin{center}
	    \includegraphics[width=.88\linewidth]{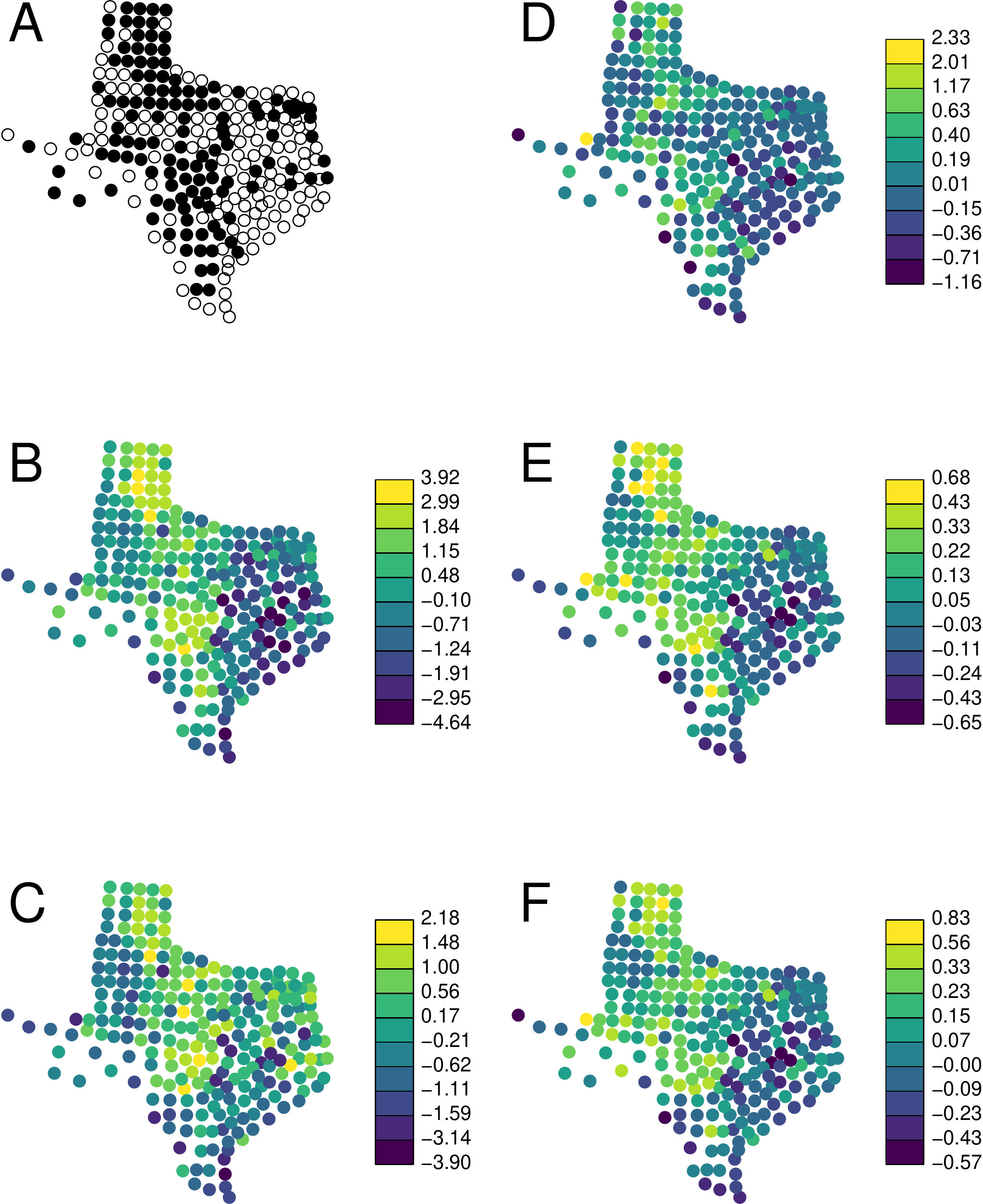}
  \end{center}
  \caption{Raw data and predicted spatial random effects ($\bw$) for the Texas turnout data. A) raw binary data, where open circles are zeros and solid circles are ones, B) predicted $\hat{\bw}$ using SAR model for binary data, C) predicted $\hat{\bw}$ using CAR model for binary data, D) logit-transformed proportional turnout data, E) predicted $\hat{\bw}$ using SAR model for proportional turnout data, F) predicted $\hat{\bw}$ using CAR model for proportional turnout data. \label{Fig:TexTurn_maps}}
\end{figure}

The predicted $\hat{\bw}$ values are shown in Figure~\ref{Fig:TexTurn_maps}.  A spatial visualization of the binary data shows some apparent clustering of 1's and 0's (Figure~\ref{Fig:TexTurn_maps}A).  The predicted $\hat{\bw}$ values for the binary data using a SAR model have the highest values in the northern Texas ``pan-handle'' and in central Texas (Figure~\ref{Fig:TexTurn_maps}B), and the pattern is similar for the predicted $\hat{\bw}$ values for the binary data when using a CAR model (Figure~\ref{Fig:TexTurn_maps}C).  A logit transformation of the raw proportional turnout data are shown in Figure~\ref{Fig:TexTurn_maps}D and the predicted $\hat{\bw}$ values for the SAR model (Figure~\ref{Fig:TexTurn_maps}E) appear to smooth the raw data, with a similar spatial pattern to the binary data (Figure~\ref{Fig:TexTurn_maps}B) and to the the predicted $\hat{\bw}$ values using a CAR model (Figure~\ref{Fig:TexTurn_maps}F).  Note that for spatial models $\bw$ interacts somewhat with the overall mean $\beta_{0}$.  Hence, we do not use the same breakpoints in the color ramps in Figure~\ref{Fig:TexTurn_maps}, as it is the relative patterns among subfigures that is of greatest interest.

\begin{spacing}{0.8}
\vspace{.4cm}
\noindent --------------------------------------------- \\
\textbf{Figure~\ref{Fig:TexTurn_maps} here}: Raw data and predicted spatial random effects ($\bw$) for the Texas turnout data. A) raw binary data, where open circles are zeros and solid circles are ones, B) predicted $\hat{\bw}$ using SAR model for binary data, C) predicted $\hat{\bw}$ using CAR model for binary data, D) logit-transformed proportional turnout data, E) predicted $\hat{\bw}$ using SAR model for proportional turnout data, F) predicted $\hat{\bw}$ using CAR model for proportional turnout data. \\
--------------------------------------------- \\ 
\end{spacing}

\begin{figure}[H]
  \begin{center}
	    \includegraphics[width=.95\linewidth]{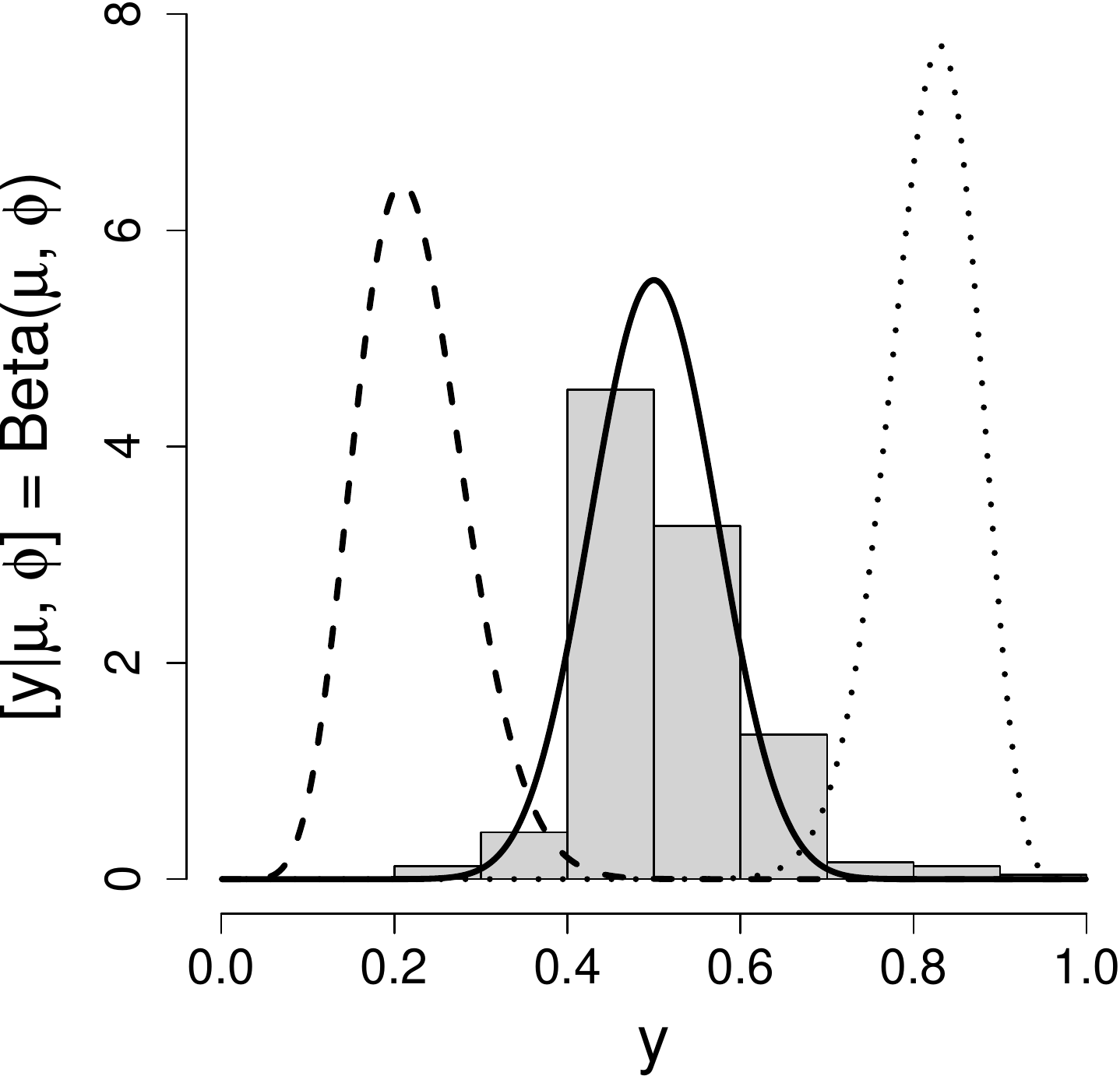}
  \end{center}
  \caption{Histogram of proportional turnout and fitted probability density functions for a beta distribution with $\phi = 46.9$ at $\mu$ values of 0.3 (dashed line), 0.5 (solid line), and 0.8 (dotted line). \label{Fig:TexTurn_histpdf}}
\end{figure}

For the beta distribution with the SAR covariance that was used for the proportional turnout data, $\phi$ was estimated to be 46.9. The density of $[y|\mu,\phi]$ will depend on $\mu$, which is affected by the explanatory variables.  In Figure~\ref{Fig:TexTurn_histpdf} we show a histogram of the raw data and the fitted probability density function, $[y|\mu, 46.9]$, which is a beta distribution, for $\mu$ = 0.3, 0.5, and 0.8.  Conditional on the mean, the fitted distributions can look quite different than the raw histogram.

\begin{spacing}{0.8}
\vspace{.4cm}
\noindent --------------------------------------------- \\
\textbf{Figure~\ref{Fig:TexTurn_histpdf} here}: Histogram of proportional turnout and fitted probability density functions for a beta distribution with $\phi = 46.9$ at $\mu$ values of 0.3 (dashed line), 0.5 (solid line), and 0.8 (dotted line). \\
--------------------------------------------- \\ 
\end{spacing}


\subsection{Harbor Seal Counts in Alaska}

For over 30 years, aerial surveys of harbor seals throughout Alaska have been flown by the Marine Mammal Laboratory of the Alaska Fisheries Science Center, part of the US government NOAA Fisheries. These surveys, which are performed primarily during the late summer months when seals are molting, are the primary method for monitoring and estimating the abundance of harbor seals (Muto et al., 2022). Based on genetic sampling, all seals in Alaska have been divided into 13 different ``stocks,'' or genetic populations.  Abundance estimates are created for each stock, and here we will use the stock known as the Sitka/Chatham Strait population. This dataset consists of 716 observations in the years 1998, 2003, 2008 - 2011, and 2015. To regulate body temperature, seals often haul out of the water, which is an how they are counted more easily than when they are in the water. All known harbor seal haul-out sites for this population were collected into 74 sample polygons. Some sample polygons were censused multiple times per year, while others were skipped in some years. For each aerial count census, explanatory variables included time-from-low-tide and time-of-day.

We use Poisson and negative binomial models in Table~\ref{tab:dDlinkDist} to formulate hierarchical GLMMs for the seal count data.  We consider a specific case of the model in \eqref{eq:lm_eta}:
\begin{linenomath*}
$$
\mathbf{w} = \bX\bbeta + \br_{1} + \bZ_{2}\br_{2} + \bepsilon,
$$
\end{linenomath*}
where $\bX$ contains a column for an overall mean; 73 columns with indicators of a mean effect for each sample polygon (as deviations from the mean for the first polygon, which was absorbed into the overall mean); a column for time-of-day, which is the elapsed fraction of the day from solar noon (the time when the sun is at the zenith); a column for time-from-low-tide, which is in hours from low tide (tide cycles last about 12 hours in this area); and columns for squared time-of-day and square time-from-low-tide.  Thus, $\bX$ has 78 columns. The $n \times 1$ random effect $\br_1$ is assumed to have the covariance structure of a stationary first-order autoregressive (AR1) time series model, so $\cov(r_{1,i}, r_{1,j}) = \sigma^{2}_{1}\rho^{|t_i - t_j|}/(1 - \rho^{2})$, where $t_i$ and $t_j$ are the years of the counts for the $i$th and $j$th observations.  Moreover, we assume that counts from distinct polygons are independent of each other, so the autocovariance only occurs among years for a given polygon, yielding a covariance matrix with a block diagonal structure. However, for some polygons there are repeated measures within year, so $\bZ_{2}$ is a design matrix created as an interaction between polygon and year, and $\sigma^{2}_{2}\bZ_{2}\bZ_{2}\upp$ allows for additional correlation among repeated samples per year within a polygon.  Not all years had repeated measures, and $\bZ_{2}$ had 518 columns with $\sigma^2_2$ as the variance of $\br_{2}$.  When using a model with a Poisson distribution, we allow for further uncorrelated overdispersion by using $\bepsilon$ as given in \eqref{eq:lm_eta}, where $\var(\bepsilon) = \sigma^2_0$, but for the negative binomial distribution, which directly allows for overdispersion, we set $\bepsilon = {\bf 0}$.
{}
\begin{table}[H] 
	\caption{Estimated fixed effects and associated quantities, under the negative binomial and Poisson models, for the harbor seal count data. The headings are the same as for Table~\ref{tab:FE_bin}.  \label{tab:FE_seals}}
\begin{center}
\begin{tabular}{|c|rrrrr|rr|}
\hline
\hline{}
{} &  \multicolumn{5}{c|}{Negative Binomial}  & \multicolumn{2}{c|}{Poisson} \\
Effect & Est. & s.e.$_{u}$ & s.e.$_{c}$  & t-val. & p-val. & s.e.$_{u}$ & s.e.$_{c}$ \\
\hline{}
time-from-low-tide & -0.081 & 0.00005 & 0.042 & 1.938 & 0.0530 & 0.041 & 0.043 \\ 
(time-from-low-tide)$^{2}$ & -0.064 & 0.00003 & 0.026 & 2.495 & 0.0128 & 0.025 & 0.026 \\ 
hour-of-day & -0.273 & 0.00013 & 0.107 & 2.542 & 0.0113 & 0.104 & 0.107 \\ 
(hour-of-day)$^{2}$ & -0.747 & 0.00017 & 0.146 & 5.120 & $<$0.0001 & 0.139 & 0.144 \\ 
\hline
\hline
\end{tabular}
\end{center}
\end{table}

For the Poisson model with AR1 covariance structure, using \eqref{eq:m2LLmargMLE} with REML we estimated $\sigma^{2}_{0} = 0.859$, $\sigma^{2}_{1} = 0.660$, $\sigma^{2}_{2} = 0.0003$ and $\rho = 0.940$, while for the negative binomial distribution with the AR1 covariance model, we estimated $\sigma^{2}_{1} = 3.637$, $\sigma^{2}_{2} = 0.0012$, $\rho = 0.997$, and $\phi = 1.529$.  The minimized value of minus twice the loglikelihood in \eqref{eq:m2LLmargMLE} was 8416.11 for the Poisson model, while it was 8242.07 for the negative binomial model, indicating that the negative binomial was a better choice. Table~\ref{tab:FE_seals} gives fixed effects estimates for the negative binomial model, except that we include the naive and corrected variances for the Poisson model as well.  It is especially interesting that when $\bepsilon$ was included in the $\bw$ values for the Poisson model, the diagonal elements of $\bC_{\bbeta}$ were large and increased only slightly when adjusted by $\mathbf{B}(-\mathbf{H}_{\mathbf{a}}^{-1})\mathbf{B}\upp$ as given in \eqref{eq:sglmm_varfe}.  This is in contrast to the negative binomial model, whose diagonal elements of $\bC_{\bbeta}$, as reflected by s.e.$_{u}$ in Table~\ref{tab:FE_seals}, are very small.  Yet, s.e.$_{c}$ is almost identical for the negative binomial and Poisson models, which is not surprising because the negative binomial and log-normal Poisson distributions have the same variance-mean relationship, differing only in the parameterization \citep[e.g.,][]{nakashima_methods_1997}.

\begin{figure}[H]
  \begin{center}
	    \includegraphics[width=.95\linewidth]{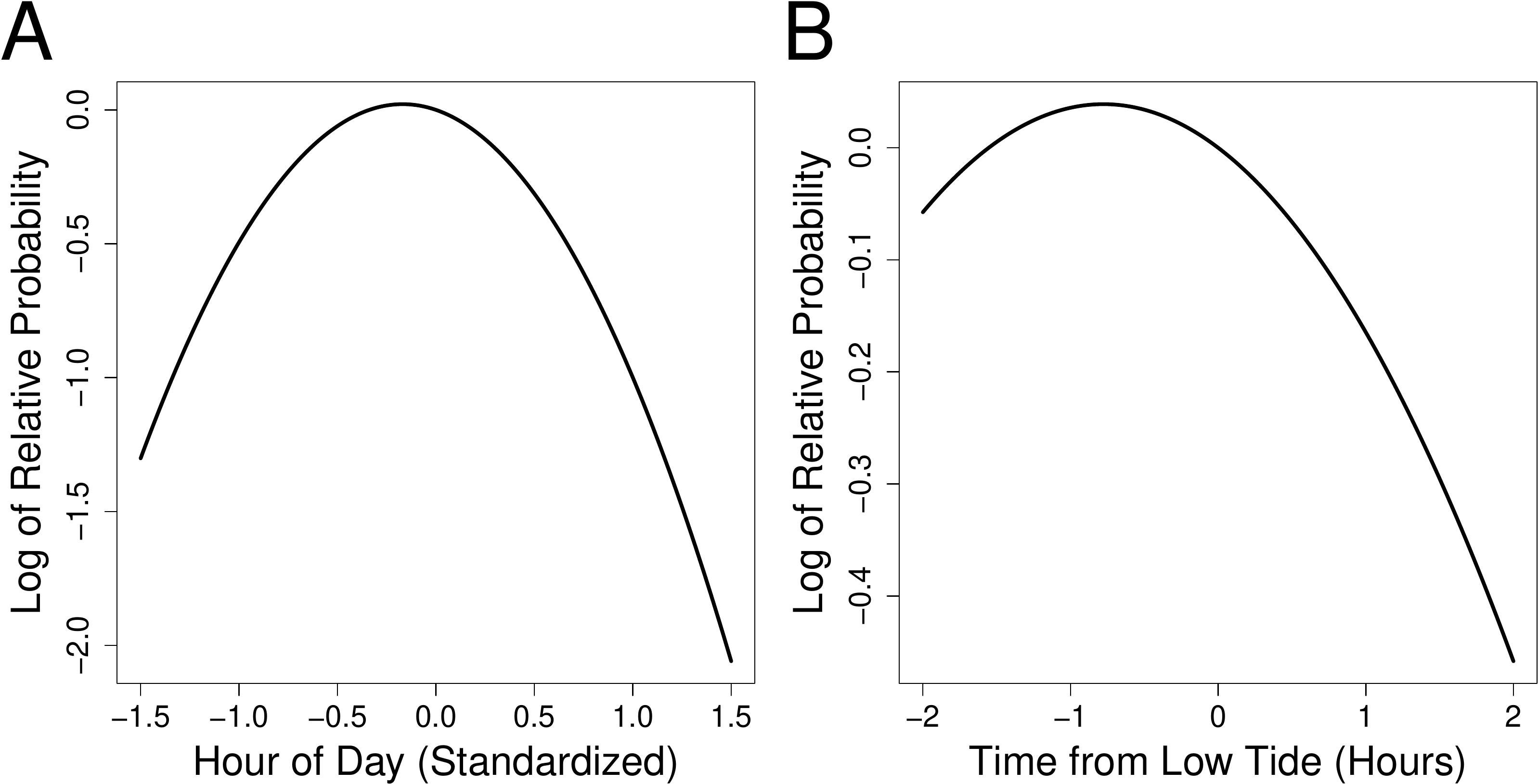}
  \end{center}
  \caption{Fitted effects of A) hour-of-day and B) time-from-low-tide on harbor seal counts.  The fitted effect shows the log of the expected proportional change when all other covariates are held at zero. \label{Fig:seals_explanvar}}
\end{figure}

The fitted explanatory variables allow insight into seal behavior, indicating that seals prefer to haul out of the water around midday and at low tides.  The model confirms that counts are highest at these times, as the coefficients in Table~\ref{tab:FE_seals} are plotted in Figure~\ref{Fig:seals_explanvar}.  This shows changes on the log scale, where all other explanatory variables are held fixed at zero, so Figure~\ref{Fig:seals_explanvar} may be interpreted as the log of the proportional change in expected counts with unit change in the explanatory variable.

\begin{spacing}{0.8}
\vspace{.4cm}
\noindent --------------------------------------------- \\
\textbf{Figure~\ref{Fig:seals_explanvar} here}: Fitted effects of A) hour-of-day and B) time-from-low-tide on harbor seal counts.  The fitted effect shows the log of the expected proportional change when all other covariates are held at zero. \\
--------------------------------------------- \\ 
\end{spacing}

\begin{figure}[H]
  \begin{center}
	    \includegraphics[width=.95\linewidth]{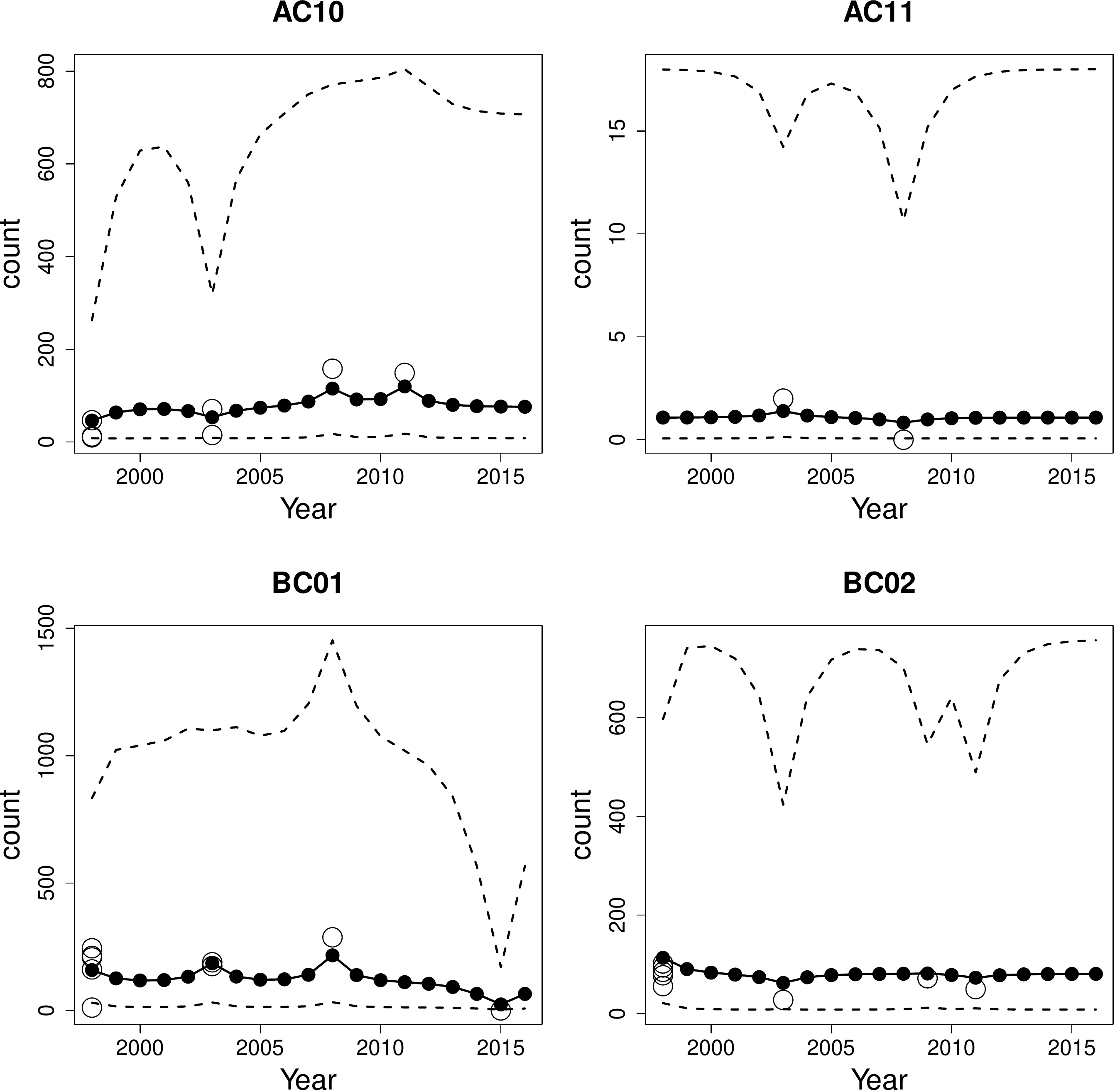}
  \end{center}
  \caption{Predicted $\exp(w)$-values for 4 of the 74 sites.  Open circles are raw counts, and solid circles are predicted $\exp(w)$-values, after back-transforming to the original scale of the data, connected by a solid line.  The dashed line shows the back-transformed prediction intervals.  \label{Fig:seals_predw}}
\end{figure}

\begin{spacing}{0.8}
\vspace{.4cm}
\noindent --------------------------------------------- \\
\textbf{Figure~\ref{Fig:seals_predw} here}: Predicted $\exp(w)$-values for 4 of the 74 sites.  Open circles are raw counts, and solid circles are predicted $\exp(w)$-values, after back-transforming to the original scale of the data, connected by a solid line.  The dashed line shows the back-transformed prediction intervals. \\
--------------------------------------------- \\ 
\end{spacing}

Our ultimate goal is to predict the abundance of seals at each site, using the results in Section~\ref{sec:pred_infer}.  Predicted $w$-values, after exponentiating, are shown in Figure~\ref{Fig:seals_predw} for 4 of the 74 different sites, which are labeled AC10, AC11, BC01, and BC02.  For each year, we predicted $\hat{\bu}$ in \eqref{eq:lm_4pred} with prediction intervals using \eqref{eq:sglmm_varpred}, and then exponentiated both predictions and prediction intervals.  We see that the errors are large, but this is not unreasonable given the relatively few observations per site.  Note that we borrowed strength across sites for estimating the autocorrelation parameter $\rho$, assuming all sites had the same amount of autocorrelation among $\bw$. Note that, in general, the predictions tend to shrink toward the overall mean for the site, but for site BC02 especially, the predictions are greater than the observed values.  This can be explained because the predictions are standardized to optimal conditions for the explanatory variables (time-of-day and time-to-low-tide). Site BC02 was counted in suboptimal conditions on almost all occasions, which is entirely possible because a high tide may occur at solar noon. It is impossible to optimize explanatory variables through a sampling design without being willing to wait and sample only when a low tide occurs at near solar noon.


\subsection{Heavy Metal Concentrations in Moss}

Cape Krusenstern National Park is in northwest Alaska, USA, and nearby is the Red Dog mine, where zinc, lead, cadmium and other heavy metals are mined. Trucks haul ore to the coast from the Red Dog Mine on a road that traverses Cape Krusenstern National Park. There is speculation that dust escapes into the environment from those trucks. Mosses obtain much of their nutrients from the air, so they are ideal biomonitors for heavy metals attached to airborne dust. In 2001 (Hasselbach et al., 2005) and again in 2006 (Neitlich et al., 2017), mosses were sampled for heavy metals, with the sampling being more dense near the road. Current annual growth of moss tissue was sampled, ground, homogenized, and then sent for laboratory analysis. Here, we just consider lead concentrations, although many other elements were analyzed. Potentially important explanatory variables that we include are distance-from-haul-road, side-of-the-road (north or south), and year of sample.  There are 365 records in the data set, with 244 from 2001 and 121 from 2006. 

Lead concentrations are inherently positive and are often skewed, which led \citet{hasselbach_spatial_2005} and \citet{neitlich_trends_2017} to transform the response to the log scale. Instead, here we use the gamma and inverse Gaussian models given in Table~\ref{tab:dDlinkDist}.  For the covariance structure, we consider a special case of the linear model \eqref{eq:lm_eta},
\begin{linenomath*}
$$
\mathbf{w} = \bX\bbeta + \br_{1} + \bZ_{2}\br_{2} + \bZ_{3}\br_{3} + \bepsilon,
$$
\end{linenomath*}
where $\bX$ contains a column for an overall mean, an indicator column for year 2006 (2001 is absorbed into the overall mean), log of distance to road (in meters), and an indicator for south of the road (north is absorbed into the overall mean).  We also consider an interaction between log distance to road and the side of the road.

The random effect $\br_{1}$ is assumed to have a geostatistical autocovariance structure given specifically by the exponential model, where $\cov(r_{1}(\bs_{i}),r_{1}(\bs_{j})) = \sigma^{2}_{1}\exp(-\delta_{i,j}/\rho)$, where $\bs_{i}$ is a vector containing the spatial coordinates of the $i$th location and $\delta_{i,j}$ is the Euclidean distance between locations $\bs_{i}$ and $\bs_{j}$. The parameter $\sigma^{2}_{1}$ is often called the partial sill, and $\rho$ is the range parameter, which controls the distance-decay rate of the autocovariance with distance. The variance of $\bepsilon$, $\sigma^{2}_{0}$, is often called the nugget effect.   We assume that responses from different years are independent.  Within year and location, at some sites, duplicate samples were obtained to account for microscale variation; that is, one handful of moss was grabbed and then another (the distance between grabs was assumed to be zero).  Hence, $\bZ_{2}$ is a design matrix with indicator variables for location; this causes increased autocorrelation for any samples from the same location.  Some samples were ground into two replicate samples for laboratory analysis, as there can be some variation in the machines that measure concentration or the way it is homogenized.  Therefore, $\bZ_{3}$ is a design matrix that contains indicator variables for a duplicate nested within location; this causes repeated replicates to have higher autocorrelation due to coming from a common duplicate.

Using the gamma distribution with the exponential covariance model and maximum likelihood (rather than REML), minus twice the log-likelihood is equal to 2318.253, and it appeared that the main effect for side-of-road was not significant. We use ML rather than REML because REML does not provide nested models for likelihood comparisons when fixed effects are changing \citep[][p. 75]{VerbekeEtAl2000LinearMixedModels}.  We refit the model without that main effect, and minus twice the log-likelihood from \eqref{eq:m2LLmargMLE} is equal to 2319.391.  Using either AIC or a likelihood ratio test, we have evidence to drop the main effect for side-of-road from the model.  This model with an interaction but no main effect of side-of-road postulates that $\mathbf{X \beta}$ at 1 m from the haul road is the same on the North and South sides.   After doing so, the marginal estimates of the covariance parameters are $\sigma^{2}_{1} = 0.1703$, $\sigma^{2}_{2} = 0.0634$, $\sigma^{2}_{3} = 0.0266$, $\sigma^{2}_{0} = 0.0023$, $\rho = 9.033$, and $\phi = 2218$.  We fit an inverse Gaussian model with the same mean and covariance structure, for which minus twice the log-likelihood is 2319.354, which is almost identical to the value for the gamma model. The covariance parameters for the inverse Gaussian model were estimated to be $\sigma^{2}_{1} = 0.1699$, $\sigma^{2}_{2} = 0.0650$, $\sigma^{2}_{3} = 0.0265$, $\sigma^{2}_{0} = 0.0028$, $\rho = 9.133$, and $\phi = 2382$, which are almost identical to their counterparts for the gamma model.  Interestingly, if we fit a normal spatial model to the log concentrations we obtain estimates $\sigma^{2}_{1} = 0.1702$, $\sigma^{2}_{2} = 0.0633$, $\sigma^{2}_{3} = 0.0267$, $\sigma^{2}_{0} = 0.0028$, $\rho = 9.055$, which are almost identical to estimates for both the gamma and inverse Gaussian models.
\begin{table}[H] 
	\caption{Estimated fixed effects table for the moss lead data when using the gamma distribution. The headings are the same as for Table~\ref{tab:FE_bin}, where here, for the p-val, we used a t-distribution with 365 - 4 = 361 degrees of freedom.    \label{tab:FE_moss}}
\begin{center}
\begin{tabular}{|c|rrrrr|}
\hline
\hline{}
Effect & Est. & s.e.$_{u}$ & s.e.$_{c}$  & t-val. & p-val.  \\
\hline{}
Intercept & 8.074 & 0.20148 & 0.20178 & 40.0169 & $<$0.0001 \\ 
Year & -0.442 & 0.22149 & 0.22157 & 1.9932 & 0.0470 \\ 
Distance to Road & -0.578 & 0.01835 & 0.01839 & 31.4377 & $<$0.0001 \\ 
Distance-to-road:Southside & -0.112 & 0.01165 & 0.01166 & 9.6119 & $<$0.0001 \\ 
\hline
\hline
\end{tabular}
\end{center}
\end{table}

\begin{figure}[H]
  \begin{center}
	    \includegraphics[width=.95\linewidth]{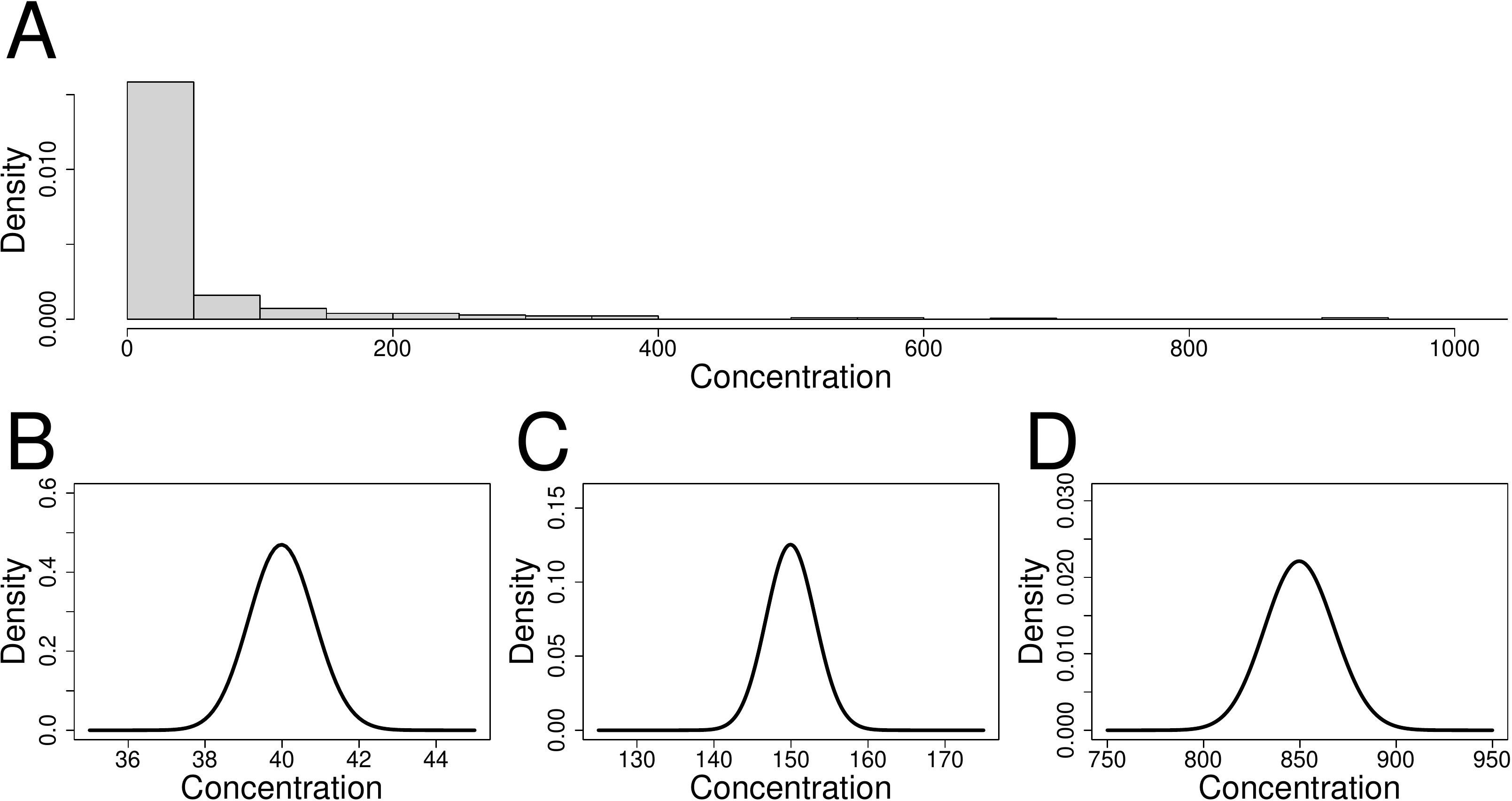}
  \end{center}
  \caption{A) Histogram of lead concentration in moss, B) fitted probability density at $\mu = 40$ with $\phi = 2218$ for the gamma distribution (solid line) C) $\mu = 150$, and D) $\mu = 850$. \label{Fig:moss_densities}}
\end{figure}

Estimates of the fixed effects tables are almost identical for all three models, so only those for the gamma model are given here; see Table~\ref{tab:FE_moss}.  Note that for this example, in contrast to the previous two examples, there is little difference in the standard errors of the estimated fixed effects based on s.e.$_{u}$ and s.e.$_{c}$.  This can happen when essentially all of the variation is captured by $\bw$ and the contribution of $\log[\by|\bg^{-1}(\bw),\bphi]$ is small in comparison.  This contribution is controlled in part by $\phi$ for the gamma and inverse Gaussian distributions, whose estimated values are very large.  To visualize, consider Figure~\ref{Fig:moss_densities}.  The histogram of the raw data (Figure~\ref{Fig:moss_densities}A) shows the extreme skewness in these data.  However, the fitted gamma distribution when the mean is 40 (conditional on the $w$-values) is quite narrow and symmetric (Figure~\ref{Fig:moss_densities}B). As the mean increases to 150 (Figure~\ref{Fig:moss_densities}C) and 850 (Figure~\ref{Fig:moss_densities}D), the gamma distributions become more dispersed, due to the fact that the variance of the gamma distribution grows as $\mu^{2}$.

\begin{spacing}{0.8}
\vspace{.4cm}
\noindent --------------------------------------------- \\
\textbf{Figure~\ref{Fig:moss_densities} here}: A) Histogram of lead concentration in moss, B) fitted probability density at $\mu = 40$ with $\phi = 2218$ for the gamma distribution (solid line) C) $\mu = 150$, and D) $\mu = 850$. \\
--------------------------------------------- \\ 
\end{spacing}

\begin{figure}[H]
  \begin{center}
	    \includegraphics[width=.95\linewidth]{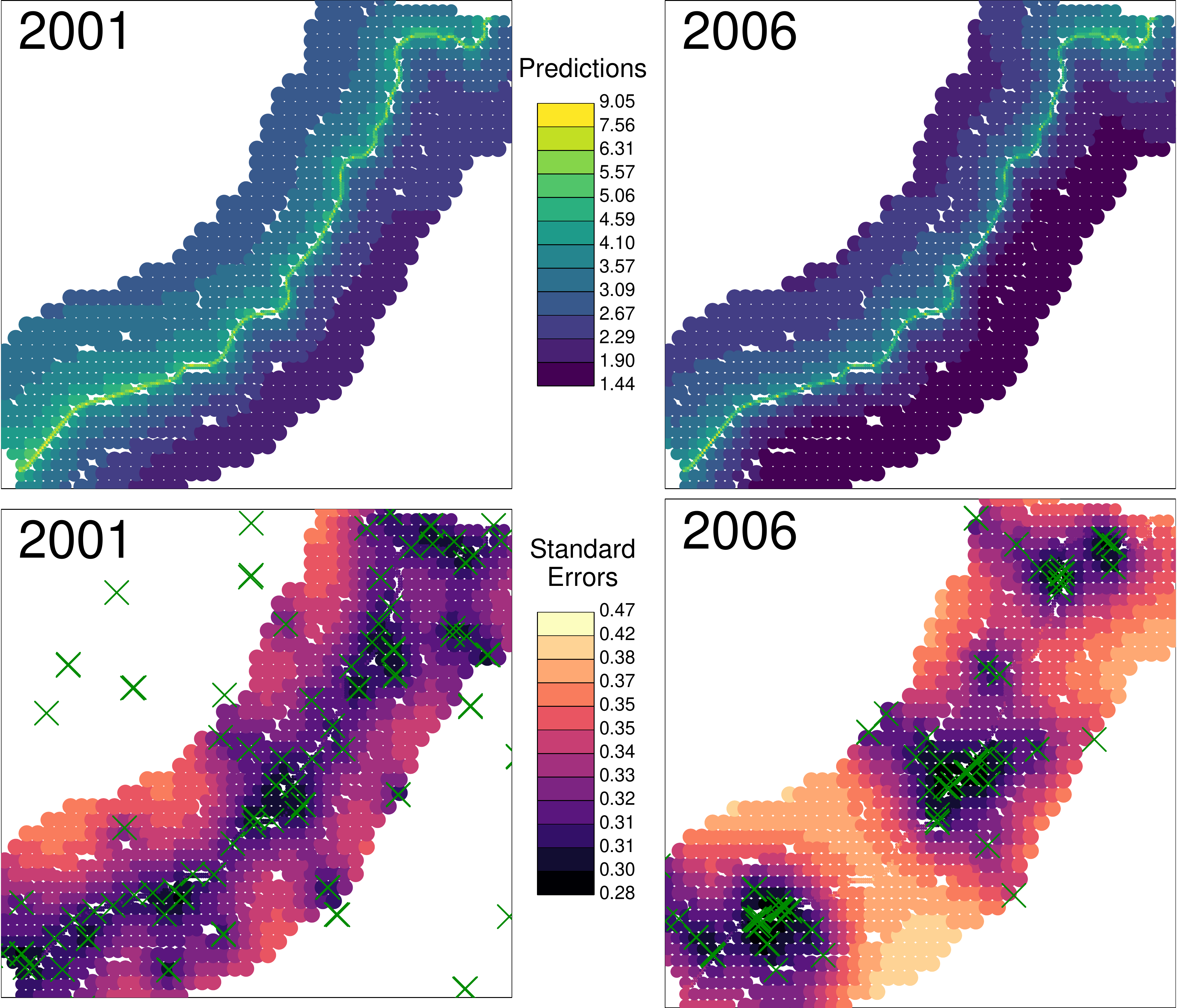}
  \end{center}
  \caption{Prediction and their standard errors for 2001 and 2006 at locations near the haul road through Cape Krusenstern National Park, Alaska.  The green $\times$ symbols show sample locations. \label{Fig:moss_maps}}
\end{figure}

Predictions of the $w$-values are similar for both the gamma and inverse Gaussian distribution models, and they are both very similar to a normal model for the log concentrations.  Predictions of $\bw$ by years 2001 and 2006 under the gamma model are shown in Figure~\ref{Fig:moss_maps}. The prediction locations are divided into three groups, one which was closely spaced near the haul road, with each successive group having more coarsely spaced locations as it gets farther from the road.   It is clear that predicted values are largest near the road (Figure~\ref{Fig:moss_maps}), but also that the predicted values generally decreased from 2001 to 2006, likely due to a change from 2001 to 2006 where coverings were used on the trucks hauling ore on the road.  The prevailing winds are from the south, and it is clear that predicted values are higher on the north side of the road.  The prediction standard errors show the typical pattern in geostatistics, i.e., they are smaller near a sample or in dense concentrations of samples (Figure~\ref{Fig:moss_maps}) than further away.

\begin{spacing}{0.8}
\vspace{.4cm}
\noindent --------------------------------------------- \\
\textbf{Figure~\ref{Fig:moss_maps} here}: Prediction and their standard errors for 2001 and 2006 at locations near the haul road through Cape Krusenstern National Park, Alaska.  The green $\times$ symbols show sample locations. \\
--------------------------------------------- \\ 
\end{spacing}


\section{Discussion and Conclusions}

We have developed a very flexible framework for modeling binary, count, positive continuous, and other data types in a hierarchical generalized linear mixed model framework.  Virtually any data type can be accommodated by the many distributions that are known in statistics, and these distributions can be matched to virtually any patterned covariance matrix, where a short list is given in Table~\ref{tab:dDlinkDist}. Our examples illustrate all of the distributions in Table~\ref{tab:dDlinkDist}, and, for covariance matrices, we used CAR and SAR spatial autoregressive models, AR1 time series models, and exponential geostatistical models.  We also showed how complex covariance matrices can be created by mixing random effects with other covariance structures.  Any covariance matrix is possible in the HGLMM framework, including spatio-temporal, covariances for data on a sphere, covariances derived for linear networks such as streams and roads, etc.  Using the Laplace approximation, the resulting log-likelihood is composed of the log-likelihood of the data distribution, the ML or REML log-likelihood for normally-distributed data, and the determinant of a Hessian matrix \eqref{eq:m2LLmargMLE}.

We have developed marginal inference for three of the most common objectives in linear models.  First, in order to estimate fixed effects and make predictions, we must estimate all covariance parameters, which is accomplished from \eqref{eq:m2LLmargMLE}.  Then, it is necessary to adjust variances for the fact that $\bw$ is latent in the model, and not observed, which is accomplished from \eqref{eq:sglmm_varfe} when estimating fixed effects and from \eqref{eq:sglmm_varpred} when predicting at unsampled locations.

The models can be computationally demanding, as they require computing the determinant of the Hessian matrix and, in our implementation, its inverse as well.  Optimizing the likelihood is doubly iterative as Newton-Raphson updates are used during likelihood optimization for covariance parameters, requiring $\bH\upi$ for each update.  While this may limit the size of data sets for our HGLMM framework, we would like to point out some time-saving features.  First, note from \eqref{eq:Hdef} that
\begin{linenomath*}
$$
\bH = [\bD_{\boldsymbol{\phi}} - \bSigma_{\btheta}^{-1}] + [\bSigma_{\btheta}^{-1}\mathbf{X}](\mathbf{X}\upp\bSigma_{\btheta}^{-1}\mathbf{X})^{-1}[\mathbf{X}\upp\bSigma_{\btheta}^{-1}],
$$
\end{linenomath*} 
where we add brackets to show that this has Sherman-Morrison-Woodbury form $\bA + \bB\bC\bB\upp$ \citep{ShermanEtAl1949Adjustmentinversematrix621,Woodbury1950Invertingmodifiedmatrices}.  If $\bA$ is $n \times n$ but has a fast inverse, and $\bC$ has small dimension, then the inverse $(\bA + \bB\bC\bB\upp)\upi$ can be made much faster than a full $n \times n$ inverse.  For example, consider our second example on harbor seals with 716 records at 74 sample sites.  We assumed a time series model within site, but independence between sites, giving the covariance matrix a block diagonal structure.  Thus, $\bSigma_{\btheta}^{-1}$ has a block-wise inverse, and $\bD_{\boldsymbol{\phi}}$ is diagonal, so $[\bD_{\boldsymbol{\phi}} - \bSigma_{\btheta}^{-1}]\upi$ can be inverted block-wise, which is much faster than a single inverse for the whole $n \times n$ matrix.  Similarly, $\bZ_{k}\bZ_{k}\upp$ is often block-diagonal, and multiple variance components can use the Sherman-Morrison-Woodbury theorem recursively \citep{dumelle_linear_2021}.

An important consideration for these models is the interplay of the independent component $\bepsilon$ in \eqref{eq:lm_eta} and $\bphi$ in $[\by|\bg^{-1}(\bw),\bphi]$. The parameters in $\bphi$ often control variance, and can be confounded with $\bepsilon$.  As an extreme example, suppose that $[\by|\bg^{-1}(\bw),\bphi]$ is a normal distribution where $\bg^{-1}(\bw)$ has the identity function for each element and \ $\phi$ has but one element -- the variance parameter.  Then $\phi$ and $\sigma_{0}^{2}$ will not be identifiable.  More often, $\bphi$ controls how variance is related to the mean, but we expect that there still can be some confounding.  For any particular data set, this can be investigated though log-likelihood plots of $\sigma_{0}^{2}$ and $\phi$, similar to Figure~\ref{Fig:sglm_likelihood_estimation}C, or with more experience on how these parameters interact for particular models.

The HGLMM framework in this paper can be contrasted to the mixed model extension of GLMs.  The GLM framework is inspired by the regular exponential family of distributions, and these lead to what are called the ``canonical'' link functions.  For example, the canonical link function for the gamma distribution is $-1/\mu$, but it is often changed to $w = g(\mu) = 1/\mu$.  However, that implies that $\mu = g\upi(w) = 1/w$, but because $w$ can be negative, it is possible for $\mu$ to have negative values.  In a moment-based modeling framework using pseudo-likelihood with iteratively weighted least squares, this can be tolerated if the values stay fairly close to the parameter space, and it allows for a wide variety of link functions, which provides a great amount of flexibility.  However, in the HGLMM framework, which is fully parametric, the evaluation of the log-likelihood for $[\by|\bg^{-1}(\bw),\bphi]$ is not possible if $\bg^{-1}(\bw)$ is outside of the parameter space for the mean.  For HGLMMs, link and mean functions must be chosen to respect the parameter space.

We have given a broad outline of marginal inference under the HGLMM.  There are many topics to explore that were not mentioned.  For example, we may want to make inference on predictions where $\bw$ is back-transformed as $\bg^{-1}(\bw)$, and where the variability of $\by|\bw$ is added.  We may also desire inference for further functions of $\bg^{-1}(\bw)$ such as block averages.  Likewise, we may want inferences on random effects (best linear unbiased predictions) of $\br_{i}$ in \eqref{eq:lm_eta}.  Like most linear models, we can consider linear combinations of $\bbeta$, or contrasts of $\bbeta$ parameters, in making inferences on fitted models, treatment effects, etc.  We only covered the basic framework is this paper and there are many further research topics to develop.

\section*{Acknowledgments}
The project received financial support from the National Marine Fisheries Service, NOAA and the U.S. Environmental Protection Agency (EPA).  The findings and conclusions in the paper are those of the author(s) and do not necessarily represent the views of the reviewers nor the EPA or the National Marine Fisheries Service, NOAA. Any use of trade, product, or firm names does not imply an endorsement by the US Government.

\section*{Data and Software Availability}
All data and code are in the Github repository, \\ \texttt{ https://github.com/jayverhoef/hglmm.git}


\bibliographystyle{asa}
\bibliography{hglmm.bib}

\newpage 

\section*{FIGURES}

\clearpage
\setcounter{equation}{0}
\renewcommand{\theequation}{A.\arabic{equation}}
\setcounter{figure}{0}
\renewcommand{\thefigure}{A.\arabic{figure}}
\section{APPENDIX}


\subsection{Derivation of REML from Integration}
\noindent
Consider a multivariate normal distribution for a general linear model,
\begin{equation} \label{eq:MVNdist}
[\mathbf{y};\bbeta,\btheta] = 
\frac{\exp\left(-\frac{1}{2}(\mathbf{y} - \mathbf{X}\bbeta)\upp\bSigma^{-1}(\mathbf{y} - \mathbf{X}\bbeta)\right)}{(2\pi)^{n/2}|\bSigma|^{1/2}},
\end{equation}
where $\mathbf{y}$ is an $n \times 1$ vector for the response variable, $\mathbf{X}$ is a $n \times p$ design matrix of explanatory variables, $\bbeta$ is a $p \times 1$ vector of fixed effects, $\btheta$ contains covariance parameters contained in the $n \times n$ covariance matrix $\bSigma$. It is possible to obtain REML equations by integrating out the fixed effects $\bbeta$,
$$
\int_{\mathbb{R}^p} f(\mathbf{y};\bbeta,\btheta) d\bbeta,
$$
to obtain a likelihood that is a function of just the covariance parameters $\btheta$ and the data $\mathbf{y}$.  In particular
$$
-2 \ln \left(\int_{\mathbb{R}^p} f(\mathbf{y};\bbeta,\btheta) d\bbeta \right) = (n - p)\ln(2\pi) + \ln|\bSigma| + \ln|\mathbf{X}\upp\bSigma^{-1}\mathbf{X}| + (\mathbf{y} - \mathbf{X}\hat{\bbeta})\upp\bSigma^{-1}(\mathbf{y} - \mathbf{X}\hat{\bbeta}),
$$
where $\hat{\bbeta} = (\mathbf{X}\upp\bSigma^{-1}\mathbf{X})^{-1}\mathbf{X}\upp\bSigma^{-1}\mathbf{y}$. \\

{\large \flushleft \textbf{Proof}}

\flushleft Write \eqref{eq:MVNdist} as
\begin{align*}
[\mathbf{y};\bbeta,\btheta] & = 
\frac{\exp\left(-\frac{1}{2}(\mathbf{y} - \mathbf{X}\hat{\bbeta} + \mathbf{X}\hat{\bbeta} - \mathbf{X}\bbeta)\upp\bSigma^{-1}(\mathbf{y} - \mathbf{X}\hat{\bbeta} + \mathbf{X}\hat{\bbeta} -  \mathbf{X}\bbeta)\right)}{(2\pi)^{n/2}|\bSigma|^{1/2}}, \\
{} & = 
\frac{\exp\left(-\frac{1}{2}[(\mathbf{y} - \mathbf{X}\hat{\bbeta})\bSigma^{-1}(\mathbf{y} - \mathbf{X}\hat{\bbeta}) + (\mathbf{X}\hat{\bbeta} - \mathbf{X}\bbeta)\upp\bSigma^{-1}(\mathbf{X}\hat{\bbeta} -  \mathbf{X}\bbeta) + C\right)}{(2\pi)^{n/2}|\bSigma|^{1/2}},
\end{align*}
where $C = 2(\mathbf{y} - \mathbf{X}\hat{\bbeta})\upp\bSigma^{-1}(\mathbf{X}\hat{\bbeta} - \mathbf{X}\bbeta)$ = 0.
Factor out terms that do not contain $\bbeta$,
$$
\int_{\mathbb{R}^p} [\mathbf{y};\bbeta,\btheta] d\bbeta = M
\int_{\mathbb{R}^p} \exp\left(-\frac{1}{2}(\mathbf{X}\hat{\bbeta} - \mathbf{X}\bbeta)\upp\bSigma^{-1}(\mathbf{X}\hat{\bbeta} -  \mathbf{X}\bbeta)\right)d\bbeta,
$$
where $M = \exp[-\frac{1}{2}(\mathbf{y} - \mathbf{X}\hat{\bbeta})\bSigma^{-1}(\mathbf{y} - \mathbf{X}\hat{\bbeta})]/[(2\pi)^{n/2}|\bSigma|^{1/2}]$.
Notice that
\begin{align*}
	{} & \int_{\mathbb{R}^p} \exp\left(-\frac{1}{2}(\mathbf{X}\hat{\bbeta} - \mathbf{X}\bbeta)\upp\bSigma^{-1}(\mathbf{X}\hat{\bbeta} -  \mathbf{X}\bbeta)\right)d\bbeta, \\
  = & \int_{\mathbb{R}^p} \exp\left(-\frac{1}{2}(\bbeta - \hat{\bbeta})\upp(\mathbf{X}\upp\bSigma^{-1}\mathbf{X})(\bbeta -  \hat{\bbeta})\right)d\bbeta, \\
	= & 2\pi^{p/2}|(\mathbf{X}\upp\bSigma^{-1}\mathbf{X})|^{-1/2},
\end{align*}
by recalling that, for positive definite $\bA_{m \times m}$ and any conformable $\bx \ne \bzero$,
$$
	\int_{\mathbb{R}^m}\exp(-\mathbf{x}\upp\mathbf{A}\mathbf{x}/2) d\mathbf{x} = (2\pi)^{m/2}|\mathbf{A}|^{-1/2}.
$$
Hence, we arrive at
$$
[\mathbf{y};\btheta] = \int_{\mathbb{R}^p} [\mathbf{y};\bbeta,\btheta] d\bbeta = \frac{\exp\left(-\frac{1}{2}(\mathbf{y} - \mathbf{X}\hat{\bbeta})\upp\bSigma^{-1}(\mathbf{y} - \mathbf{X}\hat{\bbeta})\right)}{(2\pi)^{(n-p)/2}|\bSigma|^{1/2}|\mathbf{X}\upp\bSigma^{-1}\mathbf{X}|^{1/2}},
$$
and taking $-2 \ln [\mathbf{y};\btheta]$ we obtain the desired result.


\clearpage
\subsection{Distribution Parameterizations}

\subsubsection{Negative Binomial Distribution}

\noindent{} For the negative binomial, $y_{i}$ is a non-negative integer with probability density function (PDF)
$$
[y|\mu,\phi] = 
\frac{\Gamma(y + \phi)}{\Gamma(\phi)y!}
\left(
\frac{\mu}{\mu + \phi}
\right)^{y}
\left(
\frac{\phi}{\mu + \phi}
\right)^{\phi},
$$
where $0 < \mu < 1$, $0 < \phi$, $\textrm{E}(Y) = \mu$, $\var(Y) = \mu + \mu^{2}/\phi$, and $\Gamma(\cdot)$ is the gamma function.

\subsubsection{Gamma Distribution}

\noindent{} For the gamma distribution, $y_{i}$ is positive with PDF
$$
[y|\mu,\phi] = \frac{1}{\Gamma(\phi)} \left(\frac{\phi}{\mu}\right)^\phi y^{\phi - 1}\exp\left(\frac{-y\phi}{\mu}\right),
$$
where $0 < \mu$, $0 < \phi$, $\textrm{E}(Y) = \mu$, and $\var(Y) = \mu^{2}/\phi$.{}

\subsubsection{Beta Distribution}

\noindent{} For the beta distribution, $0 < y_{i} < 1$ with PDF
$$
[y|\mu,\phi] = \frac{\Gamma(\phi)}{\Gamma(\mu\phi)\Gamma((1-\mu)\phi)}y^{\mu\phi - 1}(1 - y)^{(1-\mu)\phi - 1},
$$
where $0 < \mu < 1$, $0 < \phi$, $\textrm{E}(Y) = \mu$, and $\var(Y) = \mu(1 - \mu)/(1 + \phi)$. 

\subsubsection{Inverse Gaussian Distribution}

\noindent{} The inverse Gaussian distribution is usually written as,
\begin{equation} \label{eq:IGdist}
[y;\mu,\lambda] = 
\sqrt{\frac{\lambda}{2\pi y^{3}}}\exp\left(-\frac{\lambda(y - \mu)^{2}}{2\mu^{2} y}\right),
\end{equation}
where $y > 0$, $\mu > 0$, and $\lambda > 0$. In this parameterization $\lambda$ is a shape parameter, and $\textrm{E}(Y) = \mu$ and $\var(Y) = \mu^{3}/\lambda$.  In order to keep $\mu$ positive and $w$ unconstrained in \eqref{eq:lm_eta}, we let $\bmu = \exp(\bw)$.  However, under this construction, from \eqref{eq:Dii}, we obtain
$$
D_{i,i} = \frac{(e^{w_{i}} - 2 y_{i})}{\phi e^{2 w_{i}} },
$$
and some $D_{i,i}$ can be positive whenever $e^{w_{i}} > 2y_{i}$, which can lead to $\bH$ in \eqref{eq:Hdef} being singular.  We propose an alternative parameterization.  For inverse Gaussian models, $\lambda$ is often scaled, and here we do so by taking $\phi = \lambda/\mu = \lambda/\exp(w)$, yielding a $\mu$-scaled-$\lambda$ inverse Gaussian model,
\begin{equation} \label{eq:phiIGdist}
[y;\mu,\lambda] = 
\sqrt{\frac{\phi\exp(w)}{2\pi y^{3}}}\exp\left(-\frac{\phi(y - \exp(w))^{2}}{2\exp(w) y}\right),
\end{equation}
where $\phi > 0$ and now $\var(Y) = \mu^{2}/\phi$.  Under this parameterization, we have
$$
D_{i,i} = -\frac{\phi(e^{2w_{i}} + y_{i}^{2})}{2 y e^{w_{i}} },
$$
which is always negative, and so \eqref{eq:Hdef} is always well-behaved.  Under this construction, we also have
$$
d_{i} = \phi\left(\frac{y}{2e^{w_{i}}} - \frac{ e^{w_{i}}}{2y}\right) + \frac{1}{2}.
$$


\end{document}